\documentclass[11pt]{article}
\usepackage{a4,geometry}
\usepackage{graphicx}
\usepackage{amsmath,amssymb,amsthm,mathtools}
\usepackage{paralist}
\usepackage{bm}
\usepackage{xspace}
\usepackage{url}
\usepackage{fullpage, prettyref}
\usepackage{boxedminipage}
\usepackage{wrapfig}
\usepackage{ifthen}
\usepackage{mdframed}
\usepackage{color,xcolor}
\usepackage{transparent}
\usepackage{enumitem}
\usepackage{varwidth}
\usepackage{framed}
\usepackage{cancel}

\usepackage{fullpage}
\usepackage{ifthen}
\usepackage{algorithm}
\usepackage[noend]{algpseudocode}
\algrenewcomment[1]{\(\triangleright\) {\small{\emph{\color{blue} #1}}}}
\algnewcommand{\LineComment}[1]{\State \(\triangleright\) \emph{\color{blue} #1}}
\algnewcommand{\Invariant}[1]{\State \(\triangleright\) \emph{\color{red} #1}}
\usepackage[pagebackref,letterpaper=true,colorlinks=true,pdfpagemode=none,urlcolor=blue,linkcolor=blue,citecolor=violet,pdfstartview=FitH]{hyperref}
\usepackage[nameinlink]{cleveref}

\newtheorem{theorem}{Theorem}

\newtheorem{lemma}{Lemma}
\newtheorem{claim}{Claim}

\newtheorem{definition}{Definition}

\newtheorem{remark}{Remark}

\newcommand{\ignore}[1]{}

\newcommand{\cC}{\mathcal{C}}

\newcommand{\cI}{\mathcal{I}}

\newcommand{\cM}{\mathcal{M}}

\newcommand{\cP}{\mathcal{P}}

\newcommand{\calI}{\mathcal{I}}

\newcommand{\calM}{\mathcal{M}}

\newcommand{\calP}{\mathcal{P}}

\newcommand{\RR}{\mathbb{R}}

\newcommand{\ZZ}{\mathbb{Z}}

\newcommand{\be}{\mathbf{e}}

\newcommand{\bn}{\mathbf{n}}

\newcommand{\bx}{\mathbf{x}}
\newcommand{\by}{\mathbf{y}}
\newcommand{\bz}{\mathbf{z}}

\newcommand{\bone}{\mathbf{1}}
\newcommand{\bzero}{\mathbf{0}}

\newcommand{\poly}{\mathrm{poly}}

\newcommand{\eps}{\ensuremath{\varepsilon}}

\newcommand{\card}[1]{\left\vert{#1}\right\vert}

\newcommand{\Omgt}{\ensuremath{\widetilde{\Omega}}}

\newcommand{\norm}[1]{\left\lVert #1 \right\rVert}

\newcommand{\etal}{{\it et al.\,}}

\newcommand\dom{\mathbf{dom}}

\colorlet{shadecolor}{blue!10}

\renewcommand{\date}[1]{\hfill{\em \tiny#1}\smallskip \noindent}

\usepackage{times}

\def\Univ{U} 
\def\sizeUniv{N} 
\def\Q{{\mathsf{Q}}}
\def\Ans{{\mathsf{Ans}}}
\def\finpartfunc{h^*} 
\def\finprimepartfunc{h^{**}} 
\def\lbfunc{h} 
\def\finfunc{f^*} 
\def\finprimefunc{f^{**}} 
\def\slack{\left(\frac{n}{2}-g\right)} 
\def\gap{g} 
\def\gaptwo{\frac{\gap r}{4}} 
\def\suffix{\ell} 
\def\constr{\mathsf{C}} 

\def\SI{Suffix Indistinguishability~}
\newcommand{\trunc}[1]{#1_{\downarrow}}

\def\th{\frac{2r}{3}}

\def\pref{\mathsf{pref}}
\def\rank{\mathsf{rk}}
\def\capa{\mathsf{cap}}

\def\calPodd{\mathcal{P}_{\mathsf{odd}}}
\def\calPeven{\mathcal{P}_{\mathsf{even}}}
\def\calModd{\mathcal{M}_{\mathsf{odd}}}
\def\calMeven{\mathcal{M}_{\mathsf{even}}}

\def\vtau{\vec{\tau}}

\begin{document}
\title{A Polynomial Lower Bound on the Number of Rounds for Parallel \\Submodular Function Minimization  and Matroid Intersection}
\author{Deeparnab Chakrabarty\footnote{Department of Computer Science, Dartmouth College. Email:~{\tt deeparnab@dartmouth.edu}. Supported in part by NSF award CCF-2041920.} \and Yu Chen\footnote{Department of Computer and Information Science, University of Pennsylvania. Email~{\tt chenyu2@cis.upenn.edu}} \and Sanjeev Khanna\footnote{Department of Computer and Information Science, University of Pennsylvania. Email~{\tt sanjeev@cis.upenn.edu} Supported in part by NSF awards CCF-1910534,
CCF-1926872, and CCF-2045128.}}
\maketitle
\begin{abstract}
	\noindent
	Submodular function minimization (SFM) and matroid intersection are fundamental 
	discrete optimization problems with applications in many fields. It is well known that both of these can be solved making $\poly(N)$ queries to a relevant oracle (evaluation oracle for SFM and rank oracle for matroid intersection), where $N$ denotes the universe size.	However, all known polynomial query algorithms are highly adaptive, requiring at least $N$ rounds of querying the oracle. A natural question is whether these can be efficiently solved in a highly parallel manner, namely, with $\poly(N)$ queries using only poly-logarithmic rounds of adaptivity. 

	An important step towards understanding the adaptivity needed for efficient parallel SFM was taken recently in the work of Balkanski and Singer  who showed that any SFM algorithm making $\poly(N)$ queries necessarily requires $\Omega(\log N/\log \log N)$ rounds. This left open the possibility of efficient SFM algorithms in poly-logarithmic rounds. For matroid intersection, even the possibility of a constant round, $\poly(N)$ query algorithm was not hitherto ruled out. \smallskip
	
	In this work, we prove that any, possibly randomized, algorithm for submodular function minimization or matroid intersection making $\poly(N)$ queries requires\footnote{Throughout the paper, we use the usual convention of using $\Omgt(f(n))$ to denote $\Omega(f(n)/\log^c f(n))$  and using  $\tilde{O}(f(n))$ to denote $O(f(n)\cdot \log^c f(n))$, for some unspecified constant $c$} $\tilde{\Omega}\left(N^{1/3}\right)$ rounds of adaptivity. In fact, we show a polynomial lower bound on the number of rounds of adaptivity even for algorithms that make at most $2^{N^{1-\delta}}$ queries, for any constant $\delta> 0$.	Therefore, even though SFM and matroid intersection are efficiently solvable, they are not highly parallelizable in the oracle model.

\end{abstract}
\thispagestyle{empty}
\newpage
\setcounter{page}{1}
\section{Introduction}

A function $f:2^\Univ \to \ZZ$ defined over subsets of a ground set $\Univ$ of $\sizeUniv$ elements is submodular if for any two sets $A\subseteq B$ and an element $e\notin B$,
the {\em marginal} of $e$ on $A$, that is, $f(A\cup e) - f(A)$ is at least $f(B\cup e) - f(B)$.
The submodular function minimization (SFM) problem is to find a subset $S$ minimizing $f(S)$ given access to an evaluation oracle for the function that returns the function value on any specified subset. SFM is a fundamental discrete optimization problem which generalizes classic problems such as minimizing global and $s\textrm{-}t$ cuts in graphs and hypergraphs, 
and more recently has found applications in areas such as image segmentation~\cite{BoykoK2004,BoykoVZ2001,KohliKT2008} and speech analysis~\cite{IyerB2013,IyerJB2013}. 

A remarkable fact is that SFM {\em can} be solved in polynomial time with polynomially many queries to the evaluation oracle. This was first established by Gr\"otschel, Lov\'asz, and Schrijver~\cite{GrotsLS1981} using the ellipsoid method. Since then, a lot of work~\cite{Cunni1985,IwataFF2001,Schri2000,Orlin2009,IwataO2009,ChakrJK2014,LacosJ2015,LeeSW2015,ChakrLSW2017,DadushVZ2018,AxelrLS2020,Jiang2021} has been done trying to understand the query complexity of SFM. The current best known algorithms are an $O(\sizeUniv^3)$-query polynomial-time and an $O(\sizeUniv^2\log \sizeUniv)$-query exponential time algorithm by Jiang~\cite{Jiang2021} building on the works~\cite{LeeSW2015,DadushVZ2018}, an $\tilde{O}(\sizeUniv^2\log M)$-query and time algorithm by Lee, Sidford, and Wong~\cite{LeeSW2015} where $|f(S)|\leq M$ for all $S\subseteq \Univ$, and an $\tilde{O}(\sizeUniv M^2)$ query and time algorithm by Axelrod, Liu, and Sidford~\cite{AxelrLS2020} improving upon~\cite{ChakrLSW2017}.

Any SFM algorithm accesses the evaluation oracle in rounds, where the queries made in a certain round depend only on the answers to queries made in previous rounds. There is a trade-off between the number of queries (per round) made by the algorithm, and the number of rounds needed to find the answer : there is an obvious $1$-round algorithm which makes all $2^\sizeUniv$ queries. All known efficient algorithms for SFM described above are {\em highly sequential}; all of them proceed in $\Omega(\sizeUniv)$ rounds.
Can the number of rounds be substantially decreased (made poly-logarithmic in $\sizeUniv$) while still keeping the number of queries bounded by $\poly(N)$? In spirit, this is related to the $\mathbf{P}$ versus $\mathbf{NC}$ question which at a high-level asks whether problems with polynomial time algorithms be solved by poly-sized circuits with poly-logarithmic depth. From a practical standpoint, given the applications of SFM to problems involving huge data and the availability of computing infrastructure to perform parallel computation, the question of low-depth parallel SFM algorithms is timely.

A study of this question was initiated by Balkanski and Singer in~\cite{BalkaS2020} who proved that any polynomial query SFM algorithm must proceed in $\Omega(\frac{\log \sizeUniv}{\log\log \sizeUniv})$ rounds. 
This still leaves open the possibility of polynomial query poly-logarithmic round algorithms. Indeed for the related problem of submodular function {\em maximization}
subject to cardinality constraint, in a different paper~\cite{BalkaS2018}, Balkanski and Singer showed that the correct answer is indeed $\tilde{\Theta}(\log \sizeUniv)$. They proved that with polynomially many queries no constant factor approximation is possible with $o\left(\frac{\log \sizeUniv}{\log\log \sizeUniv}\right)$ rounds, while an $1/3$-approximation can be obtained in $O(\log \sizeUniv)$-rounds\footnote{This result has since been improved~\cite{BalkaRS2019,ChekuQ2019a,ChekuQ2019,EneN2019,EneNV2019,LiLV2020}; see~\Cref{sec:related-works} for details.}. Can the situation be the same for SFM? 

In this paper we answer this question in the negative.
We prove a {\bf \em polynomial} lower bound on the number of rounds needed by any polynomial query SFM algorithm.

\begin{mdframed}[backgroundcolor=gray!10,topline=false,bottomline=false,leftline=false,rightline=false] 
	\begin{theorem}\label{thm:parallel-sfm-lb}
        For any constant $\delta > 0$ and any $1 \le c \le \sizeUniv^{1-\delta}$, any possibly randomized algorithm for SFM on an $\sizeUniv$ element universe making $\leq \sizeUniv^c$ evaluation oracle queries per round and succeeding with probability $\geq 2/3$
        must have $\Omega\left(\frac{\sizeUniv^{1/3}}{(c\log \sizeUniv)^{1/3}}\right)$ rounds-of-adaptivity.
		This is true even when the range of the submodular function is $\{-\sizeUniv, -\sizeUniv+1, \ldots, \sizeUniv-1, \sizeUniv\}$, and even if the algorithm is only required to output the value 
		of the minimum.
	\end{theorem}

\end{mdframed}
We note that a polynomial lower bound on the number of rounds holds even if the algorithm is allowed to make $2^{N^{1-\delta}}$ queries per round for any $\delta > 0$, and the lower bound on the number of rounds is $\tilde{\Omega}(\sizeUniv^{1/3})$ for polynomial query algorithms. Our construction also proves lower bounds on the number of rounds required for {\em approximate} submodular function minimization. In this problem, one assumes via scaling that the function's range is in $[-1,+1]$ and the goal is to return a set whose value is within an additive $\eps$ from 
the minimum. We can prove an $\Omgt(1/\eps)$-lower bound on the number of rounds required for approximate SFM. 
The only previous work ruling out $\eps$-approximate minimizers is another work of Balkanski and Singer~\cite{BalkaS2017} who proved that {\em non-adaptive} algorithms, that is single round algorithms, cannot achieve any non-trivial approximation with polynomially many queries. \smallskip

Matroid intersection is another fundamental combinatorial optimization problem generalizing the maximum cardinality bipartite matching problem and the problem of packing spanning trees and arborescences in graphs. In matroid intersection, we are given two matroids $\cM_1 = (\Univ,\cI_1)$ and $\cM_2 = (\Univ, \cI_2)$ over the same universe, and the objective is to find the largest cardinality independent set present in both matroids. There are two standard ways to access these matroids: one is via the independence oracle which says whether a set is independent in a given matroid or not, and the other is via the rank oracle, which when queried with a subset $S$ returns the size of the largest independent subset of $S$. The rank oracle is stronger. It is known via Edmond's minimax~\cite{Edmo1970} result that matroid intersection (with access via rank oracles) is, in fact, a special case of submodular function minimization. The first algorithms for matroid intersection~\cite{AignerD1971,Lawler1975,Edmo1970} made $O(\sizeUniv^3)$ {\em independence oracle} queries, which was improved to $O(N^{2.5})$ by Cunningham~\cite{Cunni1986}. More recently, Chakrabarty \etal~\cite{ChakrLSSW2019} and Nguyen~\cite{Nguyen2019note} improved the number of queries to $\tilde{O}(N^2)$. The current record holder is a randomized algorithm by Blikstad \etal~\cite{BlikBMN2021} making $\tilde{O}(N^{9/5})$ independence queries. The best algorithm using rank oracle queries 
is in~\cite{ChakrLSSW2019} which gives an $\tilde{O}(N^{1.5})$-rank oracle query algorithm. As in the case of SFM, all these algorithms are sequential requiring $\Omega(N)$-rounds of adaptivity.

The submodular functions we construct to prove~\Cref{thm:parallel-sfm-lb} are closely related to the rank functions of nested matroids, a special kind of laminar matroids. As a result, we prove a similar result as in~\Cref{thm:parallel-sfm-lb} for matroid intersection. 
\begin{mdframed}[backgroundcolor=gray!10,topline=false,bottomline=false,leftline=false,rightline=false] 
	\begin{theorem}\label{thm:parallel-mati-lb}
		For any constant $\delta > 0$ and any $1 \le c \le \sizeUniv^{1-\delta}$, any possibly randomized algorithm for matroid intersection on an $\sizeUniv$ element universe making $\leq \sizeUniv^c$ {\em rank-oracle} queries per round and succeeding with probability $\geq 2/3$
		must have $\Omega\left(\frac{\sizeUniv^{1/3}}{(c\log \sizeUniv)^{1/3}}\right)$ rounds-of-adaptivity.
		This is true even when the two matroids are {\em nested matroids}, a special class of laminar matroids, and also when the algorithm is only required to output the value of the optimum.
	\end{theorem}
\end{mdframed}

In particular, any algorithm making polynomially many queries to the rank oracle must have $\Omgt(N^{1/3})$ rounds of adaptivity, even to figure out the size of the largest common independent set. That is, even the ``decision'' version of the question (is the largest cardinality at least some parameter $K$) needs polynomially many rounds of adaptivity. \medskip

Our results shows that in the general query model, SFM and matroid intersection cannot be solved in polynomial time in poly-logarithmic rounds, even with randomization.
This is in contrast to {\em specific} explicitly described succinct SFM and matroid intersection problems. For instance, global minimum cuts in an undirected graph is in $\mathbf{NC}$~\cite{KargeM1997}, finding minimum $s\textrm{-}t$-cuts with poly-bounded capacities is in $\mathbf{RNC}$~\cite{KarpUW1986}, and linear and graphic matroid intersection is in $\mathbf{RNC}$~\cite{NaraySV1994}. More recently, inspired by some of these special cases, Gurjar and Rathi~\cite{GurjaR2020} defined a class of submodular functions called {\em linearly representable} submodular functions and gave $\mathbf{RNC}$ algorithms for the same.

Our lower bounding submodular functions fall in a class introduced by Balkanski and Singer~\cite{BalkaS2020} which we call {\em partition submodular functions}.
Given a partition $\cP = (P_1,\ldots, P_r)$ of the universe $\Univ$, the value of a partition submodular function $f(S)$ depends only on the {\em cardinalities} of the $|S\cap P_i|$'s.
In particular, $f(S) = h(\bx)$ where $\bx$ is an $r$-dimensional non-negative integer valued vector with $\bx_i := |S\cap P_i|$, and $h$ is a discrete submodular function on a hypergrid.
Note that when $r=1$, the function $h$ is a univariate concave function, and when $r=n$ we obtain general submodular functions. Thus, partition submodular functions form a nice way of capturing the complexity of a submodular function.

The~\cite{BalkaS2020} functions are partition submodular and they prove an $\Omega(r)$-lower bound for their specific functions. As we explain in~\Cref{sec:technical-overview}, their construction idea has a bottleneck of $r = O(\log \sizeUniv)$, and thus cannot prove a polynomial lower bound. Our lower bound functions are also partition submodular, and we also prove an $\Omega(r)$ lower bound though we get $r$ to be polynomially large in the size of the universe. 
Furthermore, our partition submodular functions turn out to be closely related to ranks of nested matroids which lead to our lower bound for parallel matroid intersection.

\subsection{Related Work}\label{sec:related-works}

For parallel algorithms, the depth required for the ``decision'' version and the ``search'' version may be vastly different.
In a thought provoking paper~\cite{KarpUW1988}, Karp, Upfal and Wigderson considered this question. In particular, they proved that 
any efficient algorithm  that finds a maximum independent set in a {\em single} (even a partition) matroid with access to an {\em independence oracle} must proceed in $\Omgt(N^{1/3})$ rounds.
On the other hand, with access to a {\em rank} oracle which takes $S$ and returns $r(S)$, the size of the largest independent set in $S$, there is a simple algorithm\footnote{Order elements as $e_1, \ldots, e_\sizeUniv$ and query $r(\{e_1,\ldots, e_i\})$ for all $i$, and return the points at which the rank changes.	} which makes $\sizeUniv$ queries in a single round and finds the optimal answer.
Our lower bound shows that for matroid intersection, rank oracles also suffer a polynomial lower bound, even for the decision version of the problem. At this point, we should mention a very recent work of Ghosh, Gurjar, and Raj~\cite{GhoshGR2022} which showed that if there existed poly-logarithmic round algorithms for the (weighted) decision version for matroid intersection with rank-oracles, then in fact there exists {\em deterministic} polylogarithmic round algorithms for the {\em search} version. A similar flavor result is also present in~\cite{NaraySV1994}. Unfortunately, our result proves that polylogarithmic depth is impossible for arbitrary matroids (even nested ones), even when access is via rank oracles.

The rounds-of-adaptivity versus query complexity question has seen a lot of recent work on submodular function {\em maximization}.
As mentioned before, Balkanski and Singer~\cite{BalkaS2018} introduced this problem in the context of maximizing a non-negative monotone submodular function $f(S)$ subject to a cardinality constraint $|S|\leq k$. This captures $\mathbf{NP}$-hard problems, has a {\em sequential} greedy $(1-\frac{1}{e})$-approximation algorithm~\cite{NemhaWF1978}, and obtaining anything better requires~\cite{NemhaW1978,Vondr2013} exponentially many queries. ~\cite{BalkaS2018} showed that obtaining even an $O\left(\frac{1}{\log \sizeUniv}\right)$-approximation with polynomially many queries requires $\Omega\left(\frac{\log \sizeUniv}{\log\log \sizeUniv}\right)$ rounds, and gave an $O(\log \sizeUniv)$-round, polynomial query, $\frac{1}{3}$-approximation.
Soon afterwards, several different groups~\cite{BalkaRS2019,EneN2019,FahrMZ2019,ChekuQ2019,ChekuQ2019a,EneNV2019} gave $\left(1-\frac{1}{e} - \eps\right)$-approximation algorithms making polynomially many queries which run in $\poly(\log\sizeUniv, \frac{1}{\eps})$-rounds, even when the constraint on which $S$ to pick is made more general. More recently, Li, Liu and Vondr\'ak~\cite{LiLV2020} showed that the dependence of the number of rounds on $\eps$ (the distance from $1-1/e$) must be a polynomial.
Also related is the question of maximizing a non-negative non-monotone submodular function without any constraints. It is known that a random set gives a $\frac{1}{4}$-approximation, and a sequential ``double-greedy'' $\frac{1}{2}$-approximation was given by Buchbinder, Feldman, Naor, and Schwartz~\cite{BuchbFNS2015}, and this approximation factor is tight~\cite{FeigeMV2011}. 
Chen, Feldman, and Karabasi~\cite{ChenFK2019} gave a nice parallel version obtaining an $\left(\frac{1}{2}-\eps\right)$-approximation in $O(\frac{1}{\eps})$-rounds. 

In the continuous optimization setting, the question of understanding the ``parallel complexity'' of minimizing a non-smooth convex function was first studied by Nemirovski~\cite{Nemir1994}. In particular, the paper studied the problem of minimizing a bounded-norm convex (non-smooth) function over the unit $\ell_\infty$ ball in $\sizeUniv$-dimensions, and showed that any polynomial query (value oracle or gradient oracle) algorithm which comes $\eps$-close must have $\Omgt(\sizeUniv^{1/3}\ln(1/\eps))$ rounds of adaptivity. Nemirovski~\cite{Nemir1994} conjectured that the lower bound should be $\Omgt(N\ln(1/\eps))$, and this is still an open question. When the dependence on $\eps$ is allowed to be polynomial, then the sequential vanilla gradient descent outputs an $\eps$-minimizer in $O(1/\eps^2)$-rounds (over Euclidean unit norm balls), and the question becomes whether parallelism can help over gradient descent in some regimes of $\eps$. Duchi, Bartlett, and Wainwright~\cite{DuchiBW2012} showed an $O(\sizeUniv^{1/4}/\eps)$-query algorithm which is better than gradient-descent when $\frac{1}{\eps^2} > \sqrt{\sizeUniv}$. A matching lower bound in this regime was shown recently by Bubeck et al.~\cite{BubecJLLS2019}, and this paper also gives another algorithm which has better depth dependence in some regime of $\eps$. It is worth noting that submodular function minimization can also be thought of as minimizing the Lov\'asz extension which is a non-smooth convex function. Unfortunately, the domain of interest (the unit cube) has $\ell_2$-radius $\sqrt{\sizeUniv}$, and the above algorithms do not imply ``dimension-free'' $\eps$-additive approximations for submodular function minimization. 
Our work shows that $\Omega(1/\eps)$-rounds are needed, and it is an interesting open question whether a $\poly(\sizeUniv	,\frac{1}{\eps})$-lower bound can be shown on the number of rounds, or whether one can achieve efficient $\eps$-approximations in rounds independent of $\sizeUniv$.

The question of rounds-of-adaptivity versus query complexity has been asked for many other computational models, and also is closely related to other fields such as communication complexity and streaming. We note a few results which are related to submodular function minimization. Assadi, Chen, and Khanna~\cite{AssadCK2019} considered the problem of finding the minimum $s\textrm{-}t$-cut in an undirected graph in the streaming setting. They showed that any $p$-pass algorithm must take $\Omgt(n^2/p^5)$-space, where $n$ is the number of vertices. Their result also implied that any sub-polynomial round algorithm for the $s\textrm{-}t$-cut submodular function must make $\Omgt(n^2)$ queries; note that with $O(n^2)$ queries, the whole graph can be non-adaptively learned. Rubinstein, Schramm, and Weinberg~\cite{RubinSW2018} considered the global minimum cut function in an undirected unweighted graph, and showed that $\tilde{O}(n)$ queries suffice, and their algorithm can be made to run in $O(1)$-rounds. Subsequently, Mukhopadhyay and Nanongkai~\cite{MukhoN2020} generalized this for weighted undirected graphs and gave an $\tilde{O}(n)$ query algorithm.

\section{Technical Overview}\label{sec:technical-overview}

In this section, we give a technical overview of our approach to proving a polynomial lower bound on the rounds of adaptivity. We start by describing the Balkanski-Singer~\cite{BalkaS2020} framework for proving rounds-of-adaptivity lower bounds as it serves as a starting point for our work. Our presentation will first briefly highlight why the approach taken in~\cite{BalkaS2020} cannot yield better than a logarithmic lower bound on the rounds of adaptivity and then describe the approach we take to sidestep the logarithmic bottleneck.

\paragraph{The Lower Bound Framework.}
Balkanski and Singer~\cite{BalkaS2020} consider a class of submodular functions which we call {\bf \em partition submodular functions}. Given a partition $\cP = (P_1,\ldots, P_r)$ of the universe $\Univ$, 
a set function is partition submodular if its value at a subset $S$ depends only on the {\em cardinalities} of the number of elements it contains from each part. That is, 
$f_\cP(S) = h(|S\cap P_1|, \ldots, |S\cap P_r|)$ for some function $h$ whose domain is the set of $r$-dimensional non-negative integer vectors. The lower bound framework dictates the following three conditions on the functions $h$ and the resulting partition submodular function $f_\cP$.
\begin{enumerate}[noitemsep]
	\item[(P1)] The function $h$ is defined such that $f_\cP$ is {\em submodular}.
	\item[(P2)] The last part $P_r$ is the unique minimizer of $f_P$. We also assume $f_\cP(\emptyset) = h(0,0,\ldots, 0) = 0$, and thus $f_\cP(P_r)$ is necessarily $< 0$.
	\item[(P3)] For any $1\leq i <  r$, even if we know the identity of the parts $P_1, \ldots, P_{i-1}$, a single round of polynomially many queries tells us nothing about the 
	identity of the parts $P_{i+1}$ to $P_r$. More precisely, a random re-partitioning of the elements in $P_{i+1} \cup P_{i+2} \cup \cdots \cup P_{r}$ will, with high probability, give the same values to the polynomially many queries made in the current round.
\end{enumerate}
\noindent
(P3) is the key property for proving the lower bound. The function $h$ is fixed. Let $\calP$ be the uniform distribution over partitions with given sizes $|P_1|$ to $|P_r|$ which, along with $h$, induces a distribution over submodular functions. By Yao's lemma it suffices to show that any $(r-2)$-round deterministic algorithm making polynomially many queries
fails to find the minimizer with any non-trivial probability. (P3) implies that
after $(r-2)$ rounds of queries and obtaining their answers, the algorithm cannot distinguish between two functions $f_P$ and $f_{P'}$
where the partitions $P$ and $P'$ agree on the first $(r-2)$ parts, but $(P_{r-1}, P_r)$ and $(P'_{r-1}, P'_r)$ are random re-partitioning of the elements of $P_{r-1} \cup P_r$. Since (P2) implies
the minimizer of $f_P$ is $P_r$ and $f_{P'}$ is $P'_r$, and these will be different with high probability, any algorithm will make a mistake on one of them. The non-triviality is therefore in the construction
of the ``$h$'' functions, and in particular for how large an $r$ can one manage while maintaining  (P1), (P2), and (P3).

\paragraph{The Balkanski-Singer Approach.} For now, let us fix a {\em random} partition $P := (P_1, \ldots, P_r)$ of the universe $U$. Given a subset $S$, let $\bx:= (\bx_1, \bx_2, \ldots, \bx_r)$, where $\bx_i := |S\cap P_i|$ be its {\em signature}. 
Before we describe Balkanski and Singer's construction approach, let us understand what one needs for establishing a condition like (P3). Consider the case $i=1$, that is, the first round of queries.
(P3) requires that the answers should not leak any information about $P_2, P_3, \ldots, P_r$.

Consider a query $S$. Since the partition $P$ is random, we expect $S$'s signature $\bx$ to be random as well. More precisely, we expect $\frac{\bx_i}{|P_i|}$ to be ``roughly same'' for all $i\in [r]$.
Call such vectors {\em balanced}; we are deliberately not defining them precisely at this point. For (P3) to hold, we {\bf \em must} have that $\partial_i h(\bx)$, the marginal increase in the function upon adding an element from $P_i$, 
 is the same for all $2\leq i\leq r$ for {\bf \em balanced} vectors. Otherwise, the algorithm can distinguish between different parts.
On the other hand, the marginals cannot be same for {\em all} vectors $\bx$, as that would imply the sets $P_2$ to $P_r$ have the same value, which would violate the constraint (P2) since $P_r$ is the unique minimizer.

To orchestrate this, Balkanski and Singer use the idea of masking. All marginals $\partial_ih(\bx)$ are between $[-1,1]$.
In the first round, the masking is done via the first coordinate $\frac{\bx_1}{|P_1|}$ of the signature. At a very high level, when $\frac{\bx_1}{|P_1|}$ is ``large'', all the marginals
$\partial_i h(\bx)$, for $2\leq i\leq r$, take the value $-1$, while $\partial_1 h(\bx)$ takes the value $0$. In plain English, if any set $S$ contains a large fraction of elements from $P_1$, then {\em all} elements in $P_2 \cup \cdots \cup P_r$ have marginal $-1$;
the preponderance of these $P_1$ elements masks all the other parts outs. Therefore, at the first round, after making polynomially many queries an algorithm can only perhaps detect $P_1$, but has no information about parts $P_2$ to $P_r$.

More generally one requires this kind of property to hold {\em recursively} as the algorithm discovers $P_1, P_2,$ and so on in successive rounds. In any round $i$, if one considers a set $S$ with $\frac{|S\cap P_i|}{|P_i|}$ ``large'' for some $i$, then 
for all elements $e$ in parts $P_j$, $j > i$, the marginals are $-1$. 
In this way, they are able to maintain the property (P3). Of course, one has to be careful about what occurs when $|S\cap P_i|$'s are small, and the whole construction is rather technical, but this aspect described above is key to how they maintain indistinguishability.

\paragraph{A Logarithmic Bottleneck.} Unfortunately, this powerful masking property is also a bottleneck. One can 
argue that the above construction {\bf \em cannot} have $r = \omega(\log N)$. 
Consider the first round of queries. The Balkanski-Singer masking property asserts that
if $\frac{|S\cap P_1|}{|P_1|}$ is ``large'' then {\em all} $e\in P_2\cup \cdots \cup P_r$ give a marginal of $-1$. 
In particular, if one considers the the set $S = P_1$, then the marginal of all elements in $(P_2 \cup \cdots \cup P_r)$ to $S$ is $-1$. This, along with submodularity, implies that 
$f_\cP(\Univ) \leq f(P_1) -\left(\sum_{i=2}^r |P_i|\right)$. Since $\Univ$ is not the minimizer, this needs to be $> f_\cP(P_r)$, and since all marginals are in $[-1,+1]$, we get that
\[
|P_1| \geq |P_2| + \cdots + |P_{r-1}|
\]
That is, the first part is thus required to be bigger than the sum of the rest. And recursively, the second part is bigger than the sum of the rest. And so on. This implies\footnote{It is not easy to even orchestrate a $\Omega(\log N)$ lower bound this way. The masking functions that Balkanski-Singer constructed needs to be quite delicate to preserve submodularity, and in the end, the sets $P_1$ is in fact $r$ times bigger than the rest. This leads to their $\Omega(\log N/\log\log N)$ lower bound.} $r = O(\log N)$ and therefore the Balkanski-Singer masking idea {\bf cannot} give a polynomial lower bound.

\subsection{Ideas Behind Our Construction}
Let us again focus on the first round of queries.
In the Balkanski-Singer construction, whenever $\bx_1$ is ``large'' {\bf \em irrespective} of how the other $\bx_i$'s look like, the marginals $\partial_i h(\bx) = -1$ for $i\geq 2$. 
This strong masking property led to $|P_1|$ being much larger than the sum of the remaining parts
so as to compensate for all the negative marginals coming from the elements in the other parts.

Our approach is not to set $\partial_i h(\bx)$ depending on just $\bx_1$, but {\bf \em rather by looking at the whole suffix $\bx_2 : \bx_r$}. More precisely, if $\bx_1$ is ``large'' (say, even the whole part $P_1$), but all the rest are empty, even in that case
we want {\em all} marginals $\partial_i h(\bx)$ to be in fact $+1$. Only when (almost) {\em all} coordinates $\bx_i$ are ``large'', do we switch to $\partial_i h(\bx) = -1$ for all $i\geq 2$. Therefore, in a sense, elements in any part contribute 
a negative marginal towards the function value, only after a significant number of elements from that part have already contributed positively, thus canceling out the negative conributions. This is what allows our construction to have all parts of equal size $n = N/r$, setting the stage for a polynomial lower bound.

Although deciding a marginal depending on the suffix may sound complicated, in the end our lower bound functions are simple to describe. Indeed, all marginals are in the set $\{-1, 0, +1\}$ and thus not only do we prove a polynomial lower bound on 
exact SFM, we also prove a $O(1/\eps)$-lower bound even for $\eps$-approximate SFM. Furthermore, as we explain below, our lower bounding functions are closely connected to rank functions of {\em nested} matroids, which are a special class of laminar matroids. Therefore, we also obtain lower bounds on the rounds-of-adaptivity of polynomial query matroid intersection algorithms with rank-oracle queries.
In the rest of this subsection, we give more details on how the partition submodular functions are constructed. This discussion is still kept informal and is meant to help the reader understand the rationale behind the construction. The full formal details along with all the properties we need are deferred to~\Cref{sec:formal}, which the reader can feel free to skip to. \smallskip

For our lower bound, we construct two partition submodular functions, $f_\cP(S) = \lbfunc(\bx)$ and $\finfunc_\cP = \finpartfunc(\bx)$, where (a) the minimizer of $f_P$ is the empty set and the minimizer of $\finfunc_\cP$ is the set $P_r$ (satisfying (P2)), 
and {\em both} these functions satisfy (P3) for $1\leq i < r/2$, and furthermore, any, possibly randomized, algorithm distinguishing these functions and which uses only $o(N^{1/3}/\log^{1/3} N)$ rounds of adaptivity must make super-polynomial number of queries in some round. 
It is easier to understand the functions $\lbfunc$ and $\finpartfunc$ via their marginals.
Here are the properties we desire from these marginal functions.
\begin{itemize}

	\item (Submodularity.) Both function's  marginals should be monotonically decreasing. Thus, once $\partial_j \lbfunc(\bx)$ or $\partial_j \finpartfunc(\bx)$ becomes $-1$, they should stay $-1$ for all $\by$ ``larger'' than $\bx$.
	
	\item (Unique Minima.) The part $P_r$ should be the unique minimizer for $\finpartfunc$. This restricts how often $\partial_j\finpartfunc(\bx)$ can be $-1$ when $j\neq r$. This is in tension with the previous requirement.
	
	\item (Suffix Indistinguishability.) For $i\leq r/2$ and for any $\bx$ which is {\em $i$-balanced}, that is, $\bx_i \approx \bx_{i+1} \approx \cdots \approx \bx_r$,  we need that
	$\partial_j\finpartfunc(\bx)$ and $\partial_j\lbfunc(\bx)$ for such $\bx$'s should be the {\bf \em same} for all $i+1 \leq j \leq r$. This is what we call suffix indistinguishability.
	This would also imply $\lbfunc$ and $\finpartfunc$ would give the same values on all queried points with high probability.
\end{itemize}

At any point $\bx$, let us first describe the $r$ marginals $\partial_i \lbfunc(\bx)$ for $1\leq i\leq r$. 
As mentioned above, the marginals will be in the set $\{-1, 0, +1\}$.
It is best to think of this procedure constructively as an algorithm.  Initially, all the $r$ marginals are set to $+1$. 
Next, we select up to two coordinates $a$ and $b$ in $\{1,2,\ldots, r\}$, which depend on the query point $\bx$. Given these coordinates, 
 we decrement {\em all} marginals $a\leq i\leq r$ and {\em all} marginals $b\leq i\leq r$ by $1$. 
For instance, if $r = 6$ and we choose the coordinates $a=2$ and $b = 5$ at some $\bx$, then
the marginals $(\partial_1 \lbfunc(\bx), \ldots, \partial_6 \lbfunc(\bx))$ are $(1,0,0,0,-1,-1)$. The $5$th and $6$th coordinate decrement twice and thus go from $+1$ to $-1$, while the
$2$nd, $3$rd, and $4$th coordinate only decrement once and thus go from $+1$ to $0$. The first coordinate is never decremented in this example. Note that the vector of marginals when considered from $1$ to $r$ is always in decreasing order.

The crux of the construction is, therefore, in the choice of the $a$ and the $b$ at a certain point $\bx$. These will clearly depend on $\bx$, but how? Submodularity tells us that if we move from $\bx$ to $\by = \bx + \be_i$, then the $a$'s and the $b$'s should only {\em move left}, that is, become smaller; that would ensure decreasing marginals. This in turn implies that $a$ and $b$ should be defined by the {\bf \em suffix sums} at $\bx$. More precisely, if we decide to choose $a$ and $b$ as the coordinates which {\em maximize} some function $\phi(\cdot)$ which depends on the suffix sums $\sum_{i\geq t} \bx_i$, $t$ ranging from $1$ to $r$, 
then increasing a coordinate can only move $a$'s and $b$'s to the left. This is precisely what we do, and now the crux shifts to the choice of this function $\phi(\cdot)$.

Consider an $i$-balanced vector $\bx$. We need that when all the coordinates are ``large'', then the marginals of $\finpartfunc$ should be $-1$; otherwise, $P_r$ would not be the minimizer.
Since $\lbfunc$ and $\finpartfunc$ should be indistinguishable, the same should be true for $\lbfunc$. On the other hand, when all the coordinates are ``small'', most marginals of both function should be $+1$, otherwise $\Univ$ would be the minimizer.
In sum, when the coordinates of $\bx$ are ``large'', we should have the $a$ and $b$ to the left, close to $1$; this would make most marginals $-1$.
And when they are small, $a$ and $b$ should be towards the right; this would make most marginals $+1$.
This motivates the following rule that we formalize in the next section : we define $r$ different functions (called $\suffix_t(\bx)$ for $1\leq t\leq r$)
where the $t$th function $\suffix_t(\bx)$ is the sum of $(\bx_i -\tau)$ over all coordinates $t\leq i\leq r$ where $\tau$ is a ``threshold'' which is ``close'' to $n/2$. Here $n$ is the size of each part $|P_i|$. After taking the sum over these coordinates, we further subtract an``offset'' $\gamma$. In sum, the functions look like
$\suffix_t(\bx) := \left(\sum_{i=t}^r (\bx_i - \tau) \right) - \gamma$.

We choose $a$ (respectively $b$) to be the {\bf \em odd} (respectively, {\bf \em even}) coordinate $t$ with the largest $\suffix_t(\bx)$, {\em ignoring} them if this largest value is negative.
That is, if all odd $\ell_t(\bx)$'s are negative, $a$ is undefined; if all even $\ell_t(\bx)$'s are negative, $b$ is undefined. Note that if both $a$ and $b$ are undefined, all marginals $\partial_ih(\bx)$ are $+1$; if one of them is undefined, then the marginals $\partial_ih(\bx)$ are $\{+1,0\}$. Indeed, the function $h$ which achieves such marginals can be succinctly stated as 
\[
\lbfunc(\bx) = \norm{\bx}_1 - \Big(\max(0,~\max_{a: \textrm{odd}} \suffix_a(\bx)) + 		\max(0,~\max_{b: \textrm{even}} \suffix_b(\bx))\Big)
\]
To see why $\bx$ satisfies suffix indistinguishability, consider a balanced vector $\bx$ with $\bx_1 \approx \bx_2 \cdots \approx \bx_r$.
If all of these entries $\bx_i \gg \frac{n}{2}$ for all $i$, then note that the odd/even arg-maximizers are precisely $\{1,2\}$. Thus, the marginals $\partial_ih(\bx)$'s are $(0,-1,-1,\ldots, -1)$.
On the other hand if all $\bx_i \ll \frac{n}{2}$, then due to our choice $\tau \approx \frac{n}{2}$, all $\suffix_t(\bx)$'s will be negative, and thus $\{a,b\}$ will be ignored, implying that
the marginals $\partial_ih(\bx)$ will be $(+1,+1,\ldots, +1)$. In either case, the marginals $\partial_ih(\bx)$ for $i\geq 2$ are the same, implying \SI. 
In reality, we must allow a wiggle room of ``few standard deviations'' in the $\approx$ between the $\bx_i$'s since even a random set would exhibit such a behavior. To account for this, the same wiggle room needs to provided in the threshold $\tau$ and also in the offset $\gamma$. More precisely, we need to choose $\tau = \frac{n}{2} - \gap$, where $\gap \approx \tilde{\Theta}(\sqrt{n})$, and choose $\gamma \approx \gap\cdot r$.

Indeed the fact that this gap $g = \tilde{\Theta}(\sqrt{n})$ also is the reason why our construction cannot get better than $\sizeUniv^{1/3}$ lower bound. If we take the set $S = \Univ = P_1 \cup \cdots \cup P_r$,
that is, the signature $\bx = \bn = (n,n,\ldots, n)$, then one can evaluate $h(\bn) = \frac{n}{2} - \tilde{\Theta}(r\sqrt{n})$. 
If we want $f(\Univ) > 0$, we must have $n > \Theta(r\sqrt{n})$, implying $r = \tilde{O}(\sqrt{n})$. Since $\sizeUniv = nr$, this implies $r = \tilde{O}(N^{1/3})$.

The above was the description of the function $\lbfunc$ which is non-negative. The function 
$\finpartfunc$ is simply the function $\lbfunc$ if $\bx_r <\frac{n}{2} - \frac{g}{4}$, but if $\bx_r \geq \frac{n}{2} - \frac{g}{4}$, the $r$th coordinate has marginal $-1$ irrespective of the other $\bx_j$'s. This makes $P_r$ become the minimizer of $\finfunc_P$ with value $-\Theta(g)$.
Since we only modify the behavior of the last index in going from $\lbfunc$ to $\finpartfunc$, in the beginning few rounds $\lbfunc$ and $\finpartfunc$ behave similarly. Indeed, if $\bx_r > \frac{n}{2} - \frac{g}{4}$, then any $i$-balanced vector for $i \leq r/2$, has half the coordinates $\geq \frac{n}{2} - O(g)$.
The offset $\gamma$ is chosen such that in this case $\lbfunc$ also has marginal $-1$ for the $r$th coordinate. Thus, $\lbfunc$ and $\finpartfunc$ are indistinguishable in the first $r/2$ rounds. This, in turn, shows that if an algorithm runs for $<r/2$ rounds, then it cannot distinguish between these two functions, and therefore, cannot distinguish between the case when the minimum value is $0$ and when the minimum value is $\approx -N^{1/3}$.

We end this informal description by stating how our results also imply lower bounds for {\em approximate} SFM. Since the marginals of our functions are $\{-1,0,+1\}$, the range of the function is $[-N,N]$. If we scale by a multiplicative factor $N$, we immediately get an $\tilde{\Omega}(1/\sqrt{\eps})$-lower bound on the number of rounds needed to get an $\eps$-additive approximation. However, we can boost this by a bit. The main idea is to not have $P_r$ as the minimizer in $\finpartfunc$, but have the last $r/3$ parts together be the minimizer. This is done by simply having the last $r/3$ parts behave differently in $\finpartfunc$; and this boosts the minimum value to $\approx -\Theta(gr) \approx -N^{2/3}$. This implies an $\tilde{\Omega}(1/\eps)$-lower bound on the depth required to obtain an $\eps$-additive approximation. 

\paragraph{Connection with Matroid Ranks and Matroid Intersection.}
The above description of $\lbfunc(\bx)$ may seem a bit obscure. However, they are intimately connected to rank functions of matroids, in particular, nested matroids.
Given a universe $\Univ$, consider a nested family of subsets $\cC := (\Univ = C_1 \supseteq C_2 \supseteq \cdots \supseteq C_r)$. Furthermore, let each set $C_i$ have a ``capacity'' $\capa_i$.
Then, the following family of subsets 
$\calI_\cC := \{I\subseteq \Univ : |I\cap C_i| \leq \capa_i\}$
forms a matroid. Such matroids are called nested matroids, and they form a special class of laminar matroids. 
A nested matroid can also be described using a partition $P = (P_1, \ldots, P_r)$ where $P_r = C_r$ and $P_i := C_i \setminus C_{i+1}$ for all $1\leq i < r$, and thresholds
$\tau_r = \capa_r$ and $\tau_i := \capa_i - \capa_{i+1}$ for $1\leq i < r$. 
It is not too hard to show (see~\Cref{sec:matroids}) that the rank of the matroid is given by 
\[
\rank(S) := |S| - \max\left(0,~\max_{1\leq a \leq r} \ell_a(S)\right),~~~\textrm{where}~~~ \ell_t(S) := \sum_{i\geq t} \left(|S\cap P_i| - \tau_i\right)
\]
The reader can see the connection between these rank functions and the partition submodular functions described above.
Indeed, our partition submodular functions can be decomposed as $\rank_{\cM_1}(S) + \rank_{\cM_2}(\Univ\setminus S)$ (plus a constant) for two nested matroids $\cM_1$ and $\cM_2$.
Using Edmond's minimax relationship that the cardinality of the largest common independent set in $\cM_1$ and $\cM_2$ is precisely  the minimum value of functions as above, our lower bounds for parallel SFM also prove a lower bound of $\tilde{\Omega}(N^{1/3})$ on the rounds of adaptivity required for efficient matroid intersection, even in the presence of rank oracle queries.

\section{Description of our Lower Bound Functions}\label{sec:formal}

We begin by formally defining partition submodular functions and some properties of such functions. We then describe in detail the lower bound functions that we use in the proof of~\Cref{thm:parallel-sfm-lb}.

\subsection{Partition Submodular Functions}
Let $\Univ$ be a universe of elements and $\cP = (P_1, \ldots, P_r)$ be a partition of the elements of $\Univ$. 
Let $h:\ZZ_{\geq 0}^r \to \RR$ be a function whose domain is the $r$-dimensional non-negative integer hypergrid.
Given $(\cP,h)$, one can define a set-function $f_\cP:2^\Univ \to \RR$ as follows:
\begin{equation}\label{eq:partsfm-defn}
	f_\cP(S) = h\left(|S\cap P_1|, \ldots, |S\cap P_r|\right)
\end{equation}
In plain English, the value of $f_\cP(S)$ is a function only of the {\em number} of elements of each part that is present in $S$. We say that $f_\cP$ is induced by the partition $P$ and $h$.
A {\bf \em partition submodular function} is a submodular function which is induced by some partition $P$ and some hypergrid function $h$.

A function defined by $(P,h)$ is submodular if and only if $h$ satisfies the same decreasing marginal property as $f$.
To make this precise, let us settle on some notation. Throughout the paper, for any integer $k$, we use $[k]$ to denote the set $\{0,1,\ldots, k\}$.
First, note that the domain of $h$ is the $r$-dimensional hypergrid $[|P_1|] \times [|P_2|] \times \cdots \times [|P_r|]$. For brevity's sake, we call this $\dom(h)$.
We use boldfaced letters like $\bx,\by$ to denote points in $\dom(h)$. When we write $\bx+\by$ we imply coordinate-wise sum.
Given $i\in \{1,\ldots, r\}$, we use $\be_i$ to denote the $r$-dimensional vector having $1$ at the $i$th coordinate and $0$ everywhere else. 
The function $h$ induces $r$ different {\bf \em marginal} functions defined as 
\begin{equation}\label{eq:def-marginal}
	\text{For $1\leq i\leq r$}, ~~~ \partial_i h(\bx) := h(\bx + \be_i) - h(\bx)
\end{equation}
The domain of $\partial_i h$ is $[|P_1|] \times [|P_2|] \times \cdots \times [|P_i|-1] \times \cdots \times [|P_r|]$. 
\begin{definition}\label{def:hypergrid-submodular}
	We call a function $h:\ZZ^r \to \RR$ defined over a integer hypergrid $\dom(h)$ (hypergrid) submodular if and only if
	for every $1\leq i\leq r$, for every $\bx\in \dom(h)$ with $\bx_i < |P_i|$, and every $1 \le j \le r$, we have
	\begin{equation}\label{eq:part-submod}
		\partial_jh(\bx) \geq \partial_j h(\bx+\be_i)
	\end{equation}
	
\end{definition}

\begin{lemma}\label{lem:def-part}
	A set function $f_\cP$ defined by a partition $P$ and hypergrid function $h$ as in~\eqref{eq:partsfm-defn} is (partition) submodular if and only if
	$h$ is (hypergrid) submodular.
\end{lemma}
\begin{proof}
	Let $A\subseteq \Univ$ and let $\bx$ be the $r$-dimensional integer vector with $\bx_i := |A\cap P_i|$.
	Pick elements $e, e'\in \Univ\setminus A$. Let $e\in P_i$ and $e'\in P_j$ for $1\leq i,j\leq r$. Note that $j$ could be the same as $i$.
	Then $f_\cP$ is submodular is equivalent to $f_\cP(A + e') - f_\cP(A) \geq f_\cP(A+ e+e') - f_\cP(A+e)$, which is equivalent to \eqref{eq:part-submod}.
\end{proof}
\noindent
The following lemma shows that minima of partition submodular functions can be assumed to take all or nothing of each part.

\begin{lemma}\label{lem:all-or-nothing}
	Let $f_\cP$ be a partition submodular function induced by a partition $P = (P_1, \ldots, P_r)$ and hypergrid function $h$. Let $O$ be a maximal by inclusion minimizer of $f$.
	Then, $O\cap P_i \neq \emptyset$ implies $O\cap P_i = P_i$.
\end{lemma}
\begin{proof}
	Let $\bx \in \dom(h)$ be the vector induced by $O$, that is, $\bx_i = |O\cap P_i|$ for all $1\leq i\leq r$. For the sake of contradiction, assume $0 < \bx_i < |P_i|$.
	Let $e_1$ and $e_2$ be two arbitrary elements in $O\cap P_i$ and $P_i\setminus O$ respectively. 
	Since $O$ is the minimizer, $f_\cP(O) - f_\cP(O - e_1) \leq 0$. Now note that the LHS is precisely $\partial_i h(\bx-\be_i)$. And this is also equal to $f(O - e_1 + e_2) - f(O - e_1)$ and thus this is also $\leq 0$. 
	By submodularity, however, $f(O+e_2) - f(O) \leq f(O - e_1 + e_2) - f(O - e_1)$, and thus we obtain $f(O+e_2)\leq f(O)$ which contradicts that $O$ was an inclusion-wise maximal minimizer.
\end{proof}

\subsection{Suffix Functions}

The lower bound functions we construct are partition submodular functions defined with respect to a partition $\cP = (P_1, \ldots, P_r)$ of the universe $\Univ$ of $N$ elements into $r$ parts.
The number of parts $r$ is an odd integer whose value will be set to be $\tilde{\Theta}(\sizeUniv^{1/3})$.
Each part $P_i$ has the same size $n$, where $n$ is an even positive integer such that $nr = \sizeUniv$. 
The hypergrid submodular function $h:[n]^r \to \ZZ$ which define the partition submodular function are themselves defined using {\em suffix} functions, which  we describe below.

Let $\gap$ be an integer which is divisible by $4$ and which is $\tilde{\Theta}(\sqrt{n})$. That is, $\slack$ is ``many standard deviations'' away from $\frac{n}{2}$, and in particular, any random subset of an $n$-universe set has cardinality
within $\pm \gap$ of the expected value with all but inverse polynomial probability. 
As described in the previous informal discussion, the following linear suffix functions play a key role in the description of the marginals.
Define
\begin{equation}\label{eq:lin-con}
\text{For any $1\leq t\leq r$,}~~~	\suffix_t(\bx) := \sum_{s = t}^r \left(\bx_s -\slack\right) - \gaptwo 
\end{equation}
Given $\bx$, let $a  := a(\bx) \in [r]$ be the {\em odd}-coordinate $t\in [r]$ with the largest $\ell_t(\bx)$, breaking ties towards smaller indices in case of ties. 
Let $b := b(\bx) \in [r]$ be the {\em even}-coordinate $t\in [r]$ with the largest $\ell_t(\bx)$, breaking ties towards smaller indices in case of ties. 
We call $\{a,b\}$ the largest odd-even index of $\bx$.

Now we are ready to describe our lower bounding functions. 
First define the function $\lbfunc:[n]^{r}  \to \ZZ$ as follows
\begin{mdframed}[backgroundcolor=gray!05,topline=false,bottomline=false,leftline=false,rightline=false] 
\begin{equation}\label{eq:value}
\lbfunc(\bx) = \norm{\bx}_1 - \Big(\max(0,\suffix_a(\bx)) + 		\max(0,\suffix_b(\bx))\Big) 
\end{equation}
\end{mdframed}
The above function contains the seed of the hardness, and satisfies (P1) and (P3). However, the above function, for the precise choice of $g$ we will finally choose,  will in fact be non-negative. To obtain the lower bounding functions which treats $P_r$ specially,  we define
\begin{mdframed}[backgroundcolor=gray!05,topline=false,bottomline=false,leftline=false,rightline=false] 
	\begin{equation}\label{eq:extension}
	\finpartfunc(\bx) = \begin{cases}
	\lbfunc(\bx) & \textrm{if}~ \bx_r \leq \frac{n}{2} - \frac{g}{4} \\
	\lbfunc(\trunc{\bx}) - \left(\bx_r - \left(\frac{n}{2} - \frac{g}{4}\right)\right) & \textrm{otherwise}
\end{cases}
\textrm{where}, \trunc{\bx} := \left(\bx_1, \ldots, \bx_{r-1}, \min(\bx_r, \frac{n}{2} - \frac{g}{4})\right)
	\end{equation}
\end{mdframed}

In~\Cref{sec:submod}, for completeness sake, we give a direct proof that both the functions, $\lbfunc$ and $\finpartfunc$ are hypergrid submodular. However, as we show in~\Cref{sec:matroids}, these functions arise as 
sum of rank functions of particular nested matroids, and thus give a more principled reason why these functions are submodular. 
In~\Cref{sec:unique-minima}, we show that the function $\lbfunc$ is non-negative, while 
$\finpartfunc(0,0,\ldots, 0,n)$ attains a negative value of $-g/2$. In~\Cref{sec:suffix-ind}, we show that $i$-balanced vectors, for $i < r/2$, cannot distinguish between $\lbfunc$ and $\finpartfunc$.
This, in turn, is used in~\Cref{sec:main-thm} to prove the lower bound for parallel SFM.

\subsection{Submodularity}\label{sec:marginals}\label{sec:submod}
We first prove that $\lbfunc : [n]^{r} \to \ZZ$ is submodular, and then use this to prove that  $\finpartfunc : [n]^r \to \ZZ$ is submodular.
We need to prove
\begin{mdframed}[backgroundcolor=gray!10,topline=false,bottomline=false,leftline=false,rightline=false] 	
	\begin{lemma}\label{lem:part-submod}
		Fix $\bx$ and a coordinate $1\leq i \leq r$. Let $\by := \bx+\be_i$. Let $j$ be any arbitrary coordinate.
		Then,
		\begin{equation}
			\partial_j \lbfunc(\bx) \geq 			\partial_j \lbfunc(\by) 
		\end{equation}
	\end{lemma}
\end{mdframed}
The high-level reason why $\lbfunc$ is submodular is when one moves from $\bx$ to $\by = \bx+\be_i$, the odd-even index $\{a,b\}$ of $\by$ can only ``move to the left'', that is, become smaller.
Formally, 
		\begin{claim}\label{clm:power-clm}
	Let $\bx$ be any point and let $\by := \bx + \be_i$. Suppose $a$ is the odd coordinate $t$ with the largest $\ell_t(\bx)$ breaking ties towards smaller indices.
	Suppose $a'$ is the odd coordinate $t$ with the largest $\ell_t(\by)$ breaking ties towards smaller indices.
	If $a' \neq a$, then (i) $a' \leq i < a$,  and (ii) $\suffix_{a'}(\by) = \suffix_a(\by)$. A similar statement is true for even coordinates.
\end{claim}
\begin{proof}
	First from the definition, observation that $\suffix_t(\by) = \suffix_t(\bx)$ if $t > i$ and $\suffix_t(\by) = \suffix_t(\bx) + 1$ if $t\leq i$. 
	Thus, if $a' \neq a$, we must have that $a' \leq i < a$, establishing (i). Furthermore, since $a' < a$, we must have $\suffix_a(\bx) \geq \suffix_{a'}(\bx) + 1$ for otherwise $a'$ would've been chosen with respect to $\bx$.
	Since $\suffix_{a'}(\by) \geq \suffix_a(\by)$, again by the observation of the first line, we establish (ii).
\end{proof}

To see how the claim helps in proving~\Cref{lem:part-submod}, it is
instructive to first establish how the marginals of the function defined in~\eqref{eq:value} look like. 
To this end, define the following indicator functions.
For any $1\leq t\leq n$ and for any $1\leq i\leq n$, define 
\[
\constr_t(\bx) = \begin{cases}
	-1 & \textrm{if}~ \suffix_t(\bx) \geq 0 \\
	0 & \textrm{otherwise}
\end{cases}~~~\qquad\textrm{and}~~~\qquad \constr^i_t(\bx) = \constr_t(\bx)\cdot \bone_{\{i\geq t\}}
\]
where $\bone_{\{i\geq t\}}$ is the indicator function taking the value $1$ if $i\geq t$ and $0$ otherwise.
Using these notations, we can describe the $r$ different marginals at $\bx$ succinctly as
\begin{lemma}\label{lem:mar}
	Fix $\bx$ in the domain of $\lbfunc$. Let $\{a,b\}$ be largest odd-even index of $\bx$. Then, 	
\begin{equation}\label{eq:def-mar}
\forall 1\leq i\leq r, ~~ \partial_i\lbfunc(\bx) = 1 + \constr^i_a(\bx) + \constr^i_b(\bx) \tag{Marginals}
\end{equation}
\end{lemma}
\noindent
In plain English,  given a point $\bx$, one first finds the largest odd-even index $\{a,b\}$ of $\bx$. 
If any of these function values are negative, throw them away from consideration: the suffixes aren't large enough. Next, given a coordinate $i$, the marginal $\partial_i\lbfunc(\bx)$ depends on where $i$ lies in respect to $a$ and $b$ (if they are still in consideration).
If $i$ is smaller than both, then the marginal is $1$, if $i$ is smaller than one, then the marginal is $0$, if $i$ is greater than or equal to both, the marginal is $-1$. 
Given this understanding of how the marginals look like, it is perhaps clear why~\Cref{clm:power-clm} implies submodularity : as $\{a,b\}$ move left the 
the marginal of any coordinate $j$ can only decrease when one moves to $\by$. 

\begin{proof}[Proof of~\Cref{lem:mar}]
Fix an $\bx$ and a coordinate $i$. Let $\by = \bx + \be_i$. Let's consider $\lbfunc(\by) - \lbfunc(\bx)$ using \eqref{eq:value}, and then show it is precisely as asserted in \eqref{eq:def-mar}.
First note that we can rewrite
\begin{equation}\label{eq:reformulation}
\lbfunc(\bx) = \norm{\bx}_1 + \constr_a(\bx)\suffix_a(\bx) + \constr_b(\bx)\suffix_b(\bx)
\end{equation}
\noindent	
Consider the expression $\constr_a(\by)\suffix_a(\by) - \constr_a(\bx)\suffix_a(\bx)$.
If $i < a$, then $\suffix_a(\by) = \suffix_a(\bx)$, and thus $\constr_a(\by) = \constr_a(\bx)$, and thus the expression evaluates to $0$. If $i \geq a$, then $\suffix_a(\by) = \suffix_a(\bx) + 1$.
For the expression  to contribute anything non-zero, we must have $\suffix_a(\by) \geq 1$ implying $\suffix_a(\bx) \geq 0$, or in other words, $\constr_a(\bx) = \constr_a(\by) = -1$. And in that case, we get $\constr_a(\by)\suffix_a(\by) - \constr_a(\bx)\suffix_a(\bx) = -1$. To summarize,
\[
\constr_a(\by)\suffix_a(\by) - \constr_a(\bx)\suffix_a(\bx) = \begin{cases}
0 & \textrm{if $i< a$ or if $\suffix_a(\bx) < 0$, that is, $\constr_a(\bx) = 0$} \\
-1 & \textrm{otherwise, that is, if $i \geq a$ and $\constr_a(\bx) = -1$}
\end{cases}
\]
In other words,
\begin{equation}\label{eq:diff}
\constr_a(\by)\suffix_a(\by) - \constr_a(\bx)\suffix_a(\bx) = \constr_a^i(\bx)
\end{equation}

\noindent
Now suppose $\{a',b'\}$ are the odd-even index of $\by$. The above discussion proves the claim when $\{a',b'\} = \{a,b\}$.
Indeed, plugging~\eqref{eq:diff} into \eqref{eq:reformulation}, we get
\[
\lbfunc(\by) - \lbfunc(\bx) = \underbrace{\left(\norm{\by}_1 - \norm{\bx}_1\right)}_{=1}+ \constr_a^i(\bx)+ \constr_b^i(\bx)
\]

\noindent
A little more care is needed to take care of the case when $\{a',b'\}\neq \{a,b\}$. Suppose $a\neq a'$. Then, by~\Cref{clm:power-clm}, we get that $a' < i \leq a$ and $\suffix_{a'}(\by) = \suffix_a(\by)$. 
Thus,
$
\constr_{a'}(\by)\suffix_{a'}(\by) - \constr_a(\bx)\suffix_a(\bx) = \constr_a(\by)\suffix_a(\by) - \constr_a(\bx)\suffix_a(\bx)
$
and the proof follows as in the $a'=a$ case. The case $b'\neq b$ is similar. 
\end{proof}

\begin{proof}[Proof of~\Cref{lem:part-submod}]
	Let $\{a_1,b_1\}$ be the odd-even index of $\bx$. Let $\{a_2, b_2\}$ be the odd-even index of $\succ_\by$. From the definition of the marginals, what we need to show is
	\begin{equation}\label{eq:sub-nts}
		\constr^j_{a_1}(\bx) + 	\constr^j_{b_1}(\bx) \geq 	\constr^j_{a_2}(\by) + 	\constr^j_{b_2}(\by) 
	\end{equation}
 	We will show this term by term, and focus on $a_1, a_2$.
	For any $1\leq t \leq r$, observe that $\suffix_t(\by) \geq \suffix_t(\bx)$, and thus $\constr_t(\bx) \geq \constr_t(\by)$. Thus if $a_1=a_2$, we are done.
	
	If $a_1 \neq a_2$, then by~\Cref{clm:power-clm} $a_2 \leq i < a_1$ and $\ell_{a_2}(\by) = \ell_{a_1}(\by) \geq \suffix_{a_1}(\bx)$.
	This implies $\constr_{a_1}(\bx) \geq \constr_{a_2}(\by)$. Since $a_2 < a_1$, we get that $\bone_{\{j\geq a_2\}} \geq \bone_{\{j\geq a_1\}}$. Since $\constr$ is non-positive, 
	we get $\constr^j_{a_1}(\bx) = \bone_{\{j\geq a_1\}}\cdot \constr_{a_1}(\bx) \geq \bone_{\{j\geq a_2\}}\cdot \constr_{a_2}(\by) = \constr^j_{a_2}(\by)$.
\end{proof}

\begin{lemma}\label{lem:finpart-submod}
	The function $\finpartfunc$ as defined in~\eqref{eq:extension} is submodular 
\end{lemma}
\begin{proof}
	We recall the definition. 	
	\[
	\finpartfunc(\bx) = \begin{cases}
		\lbfunc(\bx) & \textrm{if}~ \bx_r \leq \frac{n}{2} - \frac{\gap}{4} \\
		\lbfunc(\trunc{\bx}) - \left(\bx_r - \left(\frac{n}{2} - \frac{\gap}{4}\right)\right) & \textrm{otherwise}
	\end{cases}
	~~\textrm{where}, \trunc{\bx} := \left(\bx_1, \ldots, \bx_{r-1}, \min(\bx_r,\frac{n}{2} - \frac{\gap}{4})\right)
	\]
	Observe, 
	\begin{itemize}
		\item If $j\neq r$, then $\partial_j \finpartfunc (\bx) = \partial_j \lbfunc (\trunc{\bx})$.
		\item If $j = r$, then $\partial_r\finpartfunc(\bx) = -1$ if $\bx_r \geq \frac{n}{2}-\frac{g}{4}$, else $\partial_r\finpartfunc(\bx) = \partial_r\lbfunc(\bx)$.
	\end{itemize}
	Now pick $\bx\in [n]^r$, $\by := \bx+ \be_i$. Since $\trunc{\bx}$ is coordinate wise dominated by $\trunc{\by}$, we get that if $j\neq r$, 

	\[
	\partial_j\finpartfunc(\bx) = \partial_j\lbfunc(\trunc{\bx}) ~~\underbrace{\geq}_{\text{\Cref{lem:part-submod}}}~~ \partial_j\lbfunc(\trunc{\by}) = \partial_j\finpartfunc(\by) 
	\]
	If $j=r$, then either $\by_r\geq \frac{n}{2}-\frac{\gap}{4}$ and then $\partial_r\finpartfunc(\bx) \ge \partial_r\finpartfunc(\by)$ since the RHS is $-1$ and the LHS is at least that.
	Or, both $\bx_r, \by_r < \frac{n}{2}-\frac{\gap}{4}$, and thus  
	$\partial_r\finpartfunc(\bx) = \partial_r\lbfunc(\bx) \underbrace{\geq}_{\text{\Cref{lem:part-submod}}} \partial_r\lbfunc(\by) = \partial_r\finpartfunc(\by)$ .
\end{proof}
\subsection{Minimizers}\label{sec:unique-minima}

\begin{mdframed}[backgroundcolor=gray!10,topline=false,bottomline=false,leftline=false,rightline=false] 	
\begin{lemma}\label{lem:unique-minimizer}
	 Suppose the parameters $n,\gap$ and $r$ chosen such that $5\gap r \leq n$. Let $P = (P_1, \ldots, P_r)$ be any partition with $|P_i|=n$ for all $i$.
	 Let $f_\cP$ be the partition submodular function induced by $(P;\lbfunc)$ and let $\finfunc_P$ be the partition submodular function induced by $(P;\finpartfunc)$.
	 Then, $\emptyset$ is the unique minimizer of $f_\cP$ achieving the value $0$, and\footnote{In fact, one can show $P_r$ is the unique minimizer of $\finfunc_P$, but that is not needed for the lower bound.} $\finfunc_P(P_r) \leq -\frac{\gap}{2}$. 
	\end{lemma}
\end{mdframed}
\begin{proof}
It is obvious that $f_\cP(\emptyset) = \finfunc_P(\emptyset)= \lbfunc(0,0,\ldots, 0) = 0$. 
Next, observe that 
\[
\finfunc_P(P_r) = \finpartfunc(0,0,\ldots, n) = \lbfunc\left(0,0,\ldots, 0, \frac{n}{2}-\frac{\gap}{4}\right) - \left(\frac{n}{2} + \frac{\gap}{4}\right)
\]
If we let $\bz = (0,0,\ldots, 0, \frac{n}{2} - \frac{\gap}{4})$, then just using $h(\bz)\leq \norm{\bz}_1$, we get $\finfunc_P(P_r) \leq -\frac{\gap}{2}$. Indeed, when $r \geq 3$, 
this is an equality since then $\ell_t(\bz) \leq 0$ for all $t$ and $h(\bz) = \norm{\bz}_1$.	

Next, we establish that if $5\gap r \leq n$, then the minimum value $f_\cP$ takes is indeed $0$. From~\Cref{lem:all-or-nothing}, we know that the maximal minimizer of $\lbfunc$ is a vector $\bx^*$ where $\bx^*_i \in \{0,n\}$ for $1\leq i\leq r$. Now fix an arbitrary $\bx$ with $\bx_i \in \{0,n\}$ which is different from the all zeros vector. 
We claim that $\lbfunc(\bx) > 0$, which would prove the lemma. Let the number of $i$'s with $\bx_i = n$ among the coordinates $\{1,2,\ldots, r\}$ be $k \geq 1$. 

Note that for any $t\leq r$, 
\[
\suffix_t(\bx) = \sum_{i\geq t} \left(\bx_i - \slack\right) - \gaptwo \leq  \left(k-t+1\right)\cdot \left(\frac{n}{2} + \gap\right) -\gaptwo
\]

Therefore, if $\{a,b\}$ are the odd-even index of $\bx$, we get that
these $\suffix_t$ values are at most 
$k\cdot \left(\frac{n}{2} + \gap\right) -\gaptwo$ and   $\left(k - 1\right)\cdot \left(\frac{n}{2} + \gap\right) -\gaptwo$, respectively, 
since $a$ and $b$ are distinct (and occurs when $a=1$ and $b=2$). Thus,
\[
\finpartfunc(\bx) = \lbfunc(\bx) > kn - 
\max\left(0, k \cdot \left(\frac{n}{2} + \gap\right) -\gaptwo\right) -
\max\left(0, \left(k-1\right) \cdot \left(\frac{n}{2} + \gap\right) -\gaptwo\right) 
\] 
If both the max terms in the expression for $\lbfunc$ turn out to be $0$, then since $k\geq 1$, we get $\lbfunc(\bx) > n$.
If only one of them is $0$, then we get $\lbfunc(\bx) > k\left(\frac{n}{2} - \gap\right) + \gaptwo > 0$.
Otherwise, we get that
\[
\finpartfunc(\bx) = \lbfunc(\bx) > kn - (2k-1)\cdot \left(\frac{n}{2} + \gap\right) - \frac{\gap r}{2} \underbrace{\geq}_{\text{using $k\leq r$}} \frac{n}{2} - \frac{5\gap r}{2} + \gap \underbrace{>}_{\text{if}~5\gap r \leq n} 0\qedhere
\]
\end{proof}

\subsection{Suffix Indistinguishability}\label{sec:suffix-ind}
We now establish the key property about $\lbfunc$ and $\finpartfunc$ which allows us to prove a polynomial lower bound on the rounds of adaptivity. To do so, we need a definition. 

\begin{definition}
	For $1\leq i < r$, a point $\bx \in [n]^r$ is called {\bf \em $i$-balanced} if $\bx_i - \frac{\gap}{8} \leq \bx_j \leq \bx_i + \frac{\gap}{8}$ for all $j > i$. 
\end{definition}

\SI asserts that two points $\bx$ and $\bx'$ which are $i$-balanced, have the same norm, and which agree on the first $i$ coordinates have the same function value.
More precisely,
\begin{mdframed}[backgroundcolor=gray!10,topline=false,bottomline=false,leftline=false,rightline=false] 	
	\begin{lemma}[\SI]\label{lem:suff-indis}
		Let $i < \frac{r}{2}$.
		If $\bx$ and $\bx'$ are two $i$-balanced points
		with $\bx_j = \bx'_j$ for $j\leq i$ and $\norm{\bx}_1 = \norm{\bx'}_1$, then $\finpartfunc(\bx) = \finpartfunc(\bx') = \lbfunc(\bx) = \lbfunc(\bx')$.
	\end{lemma}
\end{mdframed}
\begin{proof}
We first prove \SI for $\lbfunc$, and then show that if $i < \frac{r}{2}$, then $\lbfunc$ and $\finpartfunc$ take the same value on $i$-balanced points, which implies
\SI for $\finpartfunc$ as well (for $i<\frac{r}{2}$).

\begin{claim} \label{clm:suff-indis-lbfuc}
	Let $i\leq r-2$. If $\bx$ and $\bx'$ are two $i$-balanced points
	with $\bx_j = \bx'_j$ for $j\leq i$ and $\norm{\bx}_1 = \norm{\bx'}_1$, then $\lbfunc(\bx) = \lbfunc(\bx')$.
\end{claim}

\begin{proof}
	First note that for any $t\in \{1,2,\ldots, i+1\}$, $\suffix_t(\bx) = \suffix_{t}(\bx')$; this follows from the fact that $\norm{\bx}_1 = \norm{\bx'}_1$ 
and that $\bx$ and $\bx'$ agree on the first $i$-coordinates. 
	
	\noindent
	Case 1: $\bx_i = \bx'_i < \frac{n}{2} - \frac{7g}{8}$. Since $\bx$ and $\bx'$ are both $i$-balanced, we have $\bx_j, \bx'_j < \frac{n}{2} - \frac{7g}{8} + \frac{g}{8} = \frac{n}{2} - \frac{3g}{4}$ for all $j\geq i$.  
	This, in turn, implies that for any $t\geq i$, $\suffix_t(\bx), \suffix_t(\bx')$ are both $\leq \frac{gr}{4} - \frac{gr}{4} = 0$, since each summand in the definition~\eqref{eq:lin-con} contributes at most $\frac{g}{4}$. So the largest odd (similarly, even) indexed $\ell_t(\bx)$ is either negative in which case it contributes $0$ to $h(\bx)$, or $t\in \{1,\ldots, i+1\}$ in which case it subtracts $\ell_t(\bx) = \ell_t(\bx')$ from $\norm{\bx}_1 = \norm{\bx'}_1$. Furthermore, in the latter case, the same $t$ is the maximize for $\bx'$ as well. Therefore, in either case, 
	$\lbfunc(\bx) = \lbfunc(\bx')$.
	
	\noindent
	Case 2: $\bx_i = \bx'_i \geq \frac{n}{2} - \frac{7g}{8}$. Since $\bx$ and $\bx'$ are both $i$-balanced, we have $\bx_j, \bx'_j \geq \frac{n}{2} -g$ for all $j\geq i$.  
	Thus each term in the summands of~\eqref{eq:lin-con} is $\geq 0$.
	This, in turn implies that both the odd and the even maximizers of $\suffix_t(\bx), \suffix_t(\bx')$, lie in $\{1,2,\ldots, i+1\}$. 
	Since $\ell_t(\bx) = \ell_t(\bx')$ for all such $t$'s and $\norm{\bx}_1 = \norm{\bx'}_1$, we get
that $\lbfunc(\bx) = \lbfunc(\bx')$. 
\end{proof}
\noindent

Next, we prove that when $i$ is bounded way from $r$, for any $i$-balanced vector $\bx$, we have $\finpartfunc(\bx)=\lbfunc(\bx)$. 
This lemma is useful to prove the indistinguishability of $\finpartfunc$ and $\lbfunc$.
\begin{claim} \label{lem:igood-indis}
	If $i < \frac{r}{2}$ and $\bx$ is $i$-balanced, then $\finpartfunc(\bx)=\lbfunc(\bx)$.
\end{claim}

\begin{proof}
	If $\bx_r \le \frac{n}{2}-\frac{g}{4}$, we have $\finpartfunc(\bx)=\lbfunc(\bx)$ by definition. So we only need to consider the case when $\bx_r \ge \frac{n}{2}-\frac{g}{4}$. Let $k:=\bx_r - \left(\frac{n}{2}-\frac{g}{4}\right)$, by definition $\norm{\bx}_1 = \norm{\trunc{\bx}}_1 + k$ and $\finpartfunc(\bx)=\lbfunc(\trunc{\bx}) - k$. For any $1 \le t \le r$, we have $\suffix_t(\bx) = \suffix_t(\trunc{\bx}) + k$, which means that the odd (respectively, even) index $t$ with largest $\ell_t(\bx)$ is the same for $\ell_t(\trunc{\bx})$. That is the odd-even index $\{a,b\}$ is the same for $\bx$ and $\trunc{\bx}$.
	
	Since $\bx$ is $i$-balanced and $\bx_r \geq \frac{n}{2} - \frac{g}{4}$, we have $\bx_i \ge \frac{n}{2} - \frac{3g}{8}$, and thus, for any $j \ge i$, $\bx_j \ge \frac{n}{2} - \frac{g}{2}$. 
	Thus, all summands in \eqref{eq:lin-con} for $j\geq i$ give non-negative contribution.
	This means both $a$ and $b$ lie in $\{1,2,\dots,i+1\}$. On the other hand, both $\suffix_i(\trunc{\bx})$ and $\suffix_{i+1}(\trunc{\bx})$ are at least $(r-i-1) \frac{g}{2} - \frac{gr}{4} \ge 0$ since $i \le \frac{r}{2} - 1$. So both $\suffix_a(\trunc{\bx})$ and $\suffix_b(\trunc{\bx})$ are at least $0$, which implies that both $\suffix_a(\bx)$ and $\suffix_b(\bx)$ are at least $k$ (we only need they are $\geq 0$). Therefore, we have
	\begin{equation*}
		\finpartfunc(\bx) = \lbfunc(\trunc{\bx}) - k = \left(\norm{\trunc{\bx}}_1 - \suffix_a(\trunc{\bx}) - \suffix_b(\trunc{\bx})\right) - k = \norm{\bx}_1 - \suffix_a(\bx) - \suffix_b(\bx) = \lbfunc(\bx).\qedhere
	\end{equation*}
\end{proof}
\noindent
\Cref{clm:suff-indis-lbfuc} and~\Cref{lem:igood-indis} implies the \SI property of $\finpartfunc$ and $\lbfunc$.\end{proof}

\section{Parallel SFM Lower bound : Proof of~\Cref{thm:parallel-sfm-lb}}\label{sec:main-thm}

We now prove lower bounds on the rounds-of-adaptivity for algorithms which make $\leq \sizeUniv^c$ queries per round for some $1 \le c \le N^{1-\delta}$ where $\delta>0$ is a constant.
Let $n$ be an even integer and $g$ be an integer divisible by $4$ such that $800\sqrt{cn \log n} \geq g \geq 200\sqrt{c n \log n}$. Let $r$ be the largest odd integer such that $5gr \leq n$. Finally, let $\sizeUniv = nr$. Note that 
$g = \Theta(\sizeUniv^{1/3} (c \log \sizeUniv)^{2/3})$,
$r = \Theta\left(\frac{\sizeUniv^{1/3}}{ (c \log \sizeUniv)^{1/3}}\right)$, and $n = \Theta(\sizeUniv^{2/3} (c \log \sizeUniv)^{1/3})$. Since $c \leq \sizeUniv^{1-\delta}$, we  get $n>c \sizeUniv^{2\delta/3} > c \log \sizeUniv$ and thus $g \ge 200 c \log n$. 
\begin{remark}
It is perhaps worth reminding that we are allowing the algorithm to query $N^{N^{1-\delta}}$ sets. A reader may wonder with these many queries available won't one be able to find the minimizer by brute force even in a single round. In the ``hard functions'' we construct, the minimizer has $n \approx N^{1- \frac{\delta}{3}} \gg N^{1-\delta}$ elements. And thus $N^{N^{1-\delta}}$ queries would not be able to find the minimizer by enumeration over $\approx N^n$ sets.
\end{remark}

Let $\cP = (P_1, \ldots, P_r)$ be a random equipartition of a universe $\Univ$ of $\sizeUniv$ elements into parts of size $n$. 
Given a subset $S$, let the $r$-dimensional vector $\bx$ defined as $\bx_i := |S\cap P_i|$ be the signature of $S$ with respect to $\cP$.  We say a query $S$ is {\em $i$-balanced} with respect to $\cP$  if the associated signature $\bx$ is $i$-balanced.
We use the following simple property of a random equipartition.
\begin{lemma}
	\label{lem:i-good}
	For any integer $i \in [1,\ldots,(r-1)]$, let $P_1, P_2, ..., P_{i-1}$ be a sequence of $(i-1)$ sets each of size $n$ such that for $1 \le j \le (i-1)$, the set $P_j$ is generated by choosing uniformly at random $n$ elements from $\Univ \setminus (P_1 \cup P_2 \cup ... P_{j-1})$. 
	Let $S \subset \Univ$ be any query that is chosen with possibly complete knowledge of $P_1, P_2, ..., P_{i-1}$. Then if we extend $P_1, P_2, ..., P_{i-1}$ to a uniformly at random equipartition $(P_1, ..., P_r)$ of $U$, with probability at least $1 - 1/n^{2c+3}$, the query $S$ is $i$-balanced with respect to the partition $(P_1, P_2, ..., P_r)$; here the probability is taken over the choice of $P_i, P_{i+1}, ..., P_r$.
\end{lemma}
\begin{proof}
	Let $V = \Univ \setminus (P_1 \cup P_2 \cup ... \cup P_{i-1})$. For $i \le j \le r$, let
	$X_j$ be the random variable whose value equals $|S \cap P_j|$, and let $\mu = E[X_j] = |S \cap V|/(r-i+1) \le n$. To prove the assertion of the lemma, it is sufficient to show that  with probability at least $1 - 1/n^{2c+3}$, we have $|X_j - \mu | \le g/16$ for any $j$. 

    Note that each $X_j$ is a sum of $\card{V}$ negatively correlated $0/1$ random variables. By Chernoff bound for negatively correlated random variables \cite{DubhashiR98,ImpagliazzoK10}, the probability that $X_j$ deviates from its expectation $\mu$ by more than $g/16$ is at most $2e^{\max\{-\frac{(g/16)^2}{3\mu}, -(g/16)\}} \le 2e^{-10c\log n} \le 1/n^{2c+4}$. By taking a union bound over all $i \le j \le r$, with probability at least $1 - 1/n^{2c+3}$, we have $|X_j - \mu | \le g/16$ for all such $j$. 
\end{proof}

To prove~\Cref{thm:parallel-sfm-lb}, we use Yao's minimax lemma. The distribution over hard functions is as follows. First, we sample a random equipartition $\cP$ of the $\Univ$ into $r$ parts each of size $n$. Given $\cP$ and a subset $S$, let $f_\cP(S) := \lbfunc(\bx)$ and $\finfunc_\cP(S) := \finpartfunc(\bx)$, where $\bx$ is the signature of $S$ with respect to $\cP$. Select one of $f_\cP$ and $\finfunc_\cP$ uniformly at random. This fixes the distribution over the functions, and this distribution is offered to a deterministic algorithm. We now prove that any $s$-round deterministic algorithm with $s < \frac{r}{2}$ fails to return the correct answer with probability $> 1/3$, and this would prove~\Cref{thm:parallel-sfm-lb}. In fact, we prove that with probability $\geq 1 - 1/n$, over the random equipartition $\cP$, the deterministic algorithm cannot distinguish between $f_\cP$ and $\finfunc_\cP$, that is, the answers to all the queries made by the algorithm is the same on both functions. This means that the deterministic algorithm errs with probability $\geq \frac{1}{2}\cdot (1 - \frac{1}{n}) > \frac{1}{3}$.

	An $s$-round deterministic algorithm performs a collection of queries $\Q^{(\ell)}$ at every round $1\leq \ell \leq s$ with $|\Q^{(\ell)}|\leq \sizeUniv^c \leq n^{2c}$. Let $\Ans^{(\ell)}$ denote the answers to the queries in $\Q^{(\ell)}$.
The subsets queried in $\Q^{(\ell)}$ is a deterministic function of the answers given in $\Ans^{(1)}, \ldots, \Ans^{(\ell-1)}$. After receiving the answers to the $s$th round of queries,
that is $\Ans^{(s)}$, the algorithm must return the minimizing set $S$.
We now prove that when $\cP$ is a random equipartition of $\Univ$, then with probability $1-\frac{1}{n}$, the answers $\Ans^{(\ell)}$ given to $\Q^{(\ell)}$ are the same for $f_\cP$ and $\finfunc_\cP$, if $s < \frac{r}{2}$.

	We view the process of generating the random equipartition as a game between an adversary and the algorithm where the adversary reveals the parts one-by-one.
	Specifically, the process of generating the random equipartition will be such that at the start of any round $\ell \in [1, \ldots, s]$, the adversary has only chosen and revealed to the algorithm the parts $P_1, P_2, ..., P_{\ell-1}$, and at this stage, $P_{\ell}, P_{\ell+1}, ..., P_r$ are equally likely to be any equipartition of $\Univ \setminus (P_1 \cup P_2 \cup ... \cup P_{\ell-1})$ into $(r - \ell +1)$ parts. By the end of round $\ell$, the adversary has committed and revealed to the algorithm the part $P_{\ell}$, and the game continues with one caveat. In each round, there will be a small probability (at most $1/n^2$) with which the adversary may ``fail''. This occurs at a round $\ell$ if any query made by the algorithm {\em on or before round $\ell$} turns out to be not $\ell$-balanced 
	with respect to the sampled partition at round $\ell$.  In that case, the adversary reveals all remaining parts to the algorithm (consistent with the answers given thus far), and the game terminates in the current round $\ell$ itself with the algorithm winning the game (that is, the algorithm can distinguish between $f_\cP$ and $\finfunc_\cP$). The probability of this failure event can be bound by $s/n^2 \le 1/n$, summed over all rounds. In absence of this failure event, by~\Cref{lem:suff-indis}, we know that the answers will be the same for $f_\cP$ and $\finfunc_\cP$ at the end of the algorithm, concluding the proof. We now formally describe this process.

At the start of round $1$, the adversary samples a uniformly at random equipartition of $U$, say, $\Gamma^{(1)} = (P_1^{(1)}, P_2^{(1)}, ..., P_r^{(1)})$. The algorithm reveals its set of queries for round $1$, namely, $\Q^{(1)}$. 
The adversary answers all queries in $\Q^{(1)}$ in accordance with the partition $\Gamma^{(1)}$.
By~\Cref{lem:i-good}, since $|\Q^{(1)}| \leq n^{2c}$, 
every query in $\Q^{(1)}$ is $1$-balanced with respect to the partition $\Gamma^{(1)}$, with probability at least $1 - 1/n^3$.
If this event occurs, the adversary reveals $P_1^{(1)}$ to the algorithm, and continues to the next round.
Otherwise, the adversary reveals the entire partition $\Gamma^{(1)}$ to the algorithm and the game terminates.

At the start of round $2$, the adversary samples another uniformly at random equipartition of $U$, say, $\Gamma^{(2)} = (P_1^{(2)}, P_2^{(2)}, ..., P_r^{(2)})$ subject to the constraint $P_1^{(2)} = P_1^{(1)}$. Note that $\Gamma^{(2)}$ is a uniformly at random equipartition of $U$ since $P_1^{(1)}$ was chosen uniformly at random. The algorithm reveals its set of queries for round $2$, namely, $\Q^{(2)}$. 
Again by~\Cref{lem:i-good}, we have that
(i) every query in $\Q^{(1)}$ is $1$-balanced with respect to the partition $\Gamma^{(2)}$, with probability at least $1 - 1/n^3$, and 
(ii) every query in $\Q^{(2)}$ is $2$-balanced with respect to the partition $\Gamma^{(2)}$, with probability at least $1 - 1/n^3$.
If this event occurs, the adversary answers all queries in $\Q^{(2)}$ in accordance with the partition $\Gamma^{(2)}$, and the game proceeds to the next round. The key insight here is that by~\Cref{lem:suff-indis}, if a query $S \in \Q^{(i)}$ is $i$-balanced w.r.t. some partition $(P_1, ..., P_r)$, then the function value on the query $S$ is completely determined by $P_1, P_2, ..., P_i$ and $|S|$, and does not require knowledge of $P_{i+1}, ..., P_r$. Furthermore, the value of $f_\cP(S)$ {\em and} $\finfunc_\cP(S)$ are the same.
In other words, the function value on query $S$
remains unchanged, for both $f$ and $\finfunc$, if we replace $P := (P_1, ...,P_i, P_{i+1}, ..., P_r)$ with any other partition $P' := (P_1, ..., P_i, P'_{i+1}, .., P'_r)$ such that $S$ remains $i$-balanced with respect to $P'$. So answers to all queries in $\Q^{(1)}$ are the same under both partitions $\Gamma^{(1)}$ and $\Gamma^{(2)}$. On the other hand, if either (i) or (ii) above does not occur, the adversary terminates the game and reveals the entire partition $\Gamma^{(1)}$ to the algorithm. 

In general, if the game has successfully reached round $\ell \le s$, then at the start of round $\ell$, the adversary samples a uniformly at random equipartition of $U$, say, $\Gamma^{(\ell)} = (P_1^{(\ell)}, P_2^{(\ell)}, ..., P_r^{(\ell)})$ subject to the constraints $P_1^{(\ell)} = P_1^{(1)}, P_2^{(\ell)} = P_2^{(2)}, ..., P_{\ell-1}^{(\ell)} = P_{\ell-1}^{(\ell-1)}$. Once again, note that $\Gamma^{(\ell)}$ is a uniformly at random equipartition of $U$ since $P_1^{(1)}$ was chosen uniformly at random, $P_2^{(2)}$ was chosen uniformly at random having fixed $P_1^{(1)}$, and so on. 
The algorithm now reveals its set of queries for round $\ell$, namely, $\Q^{(\ell)}$. 
By~\Cref{lem:i-good}, we have that for any fixed $i \in [1,\ldots,\ell]$, all queries in $\Q^{(i)}$ are $i$-balanced with respect to the partition $\Gamma^{(\ell)}$ with probability at least $1 - 1/n^3$ each. Thus with probability at least $1 - \ell/n^3$, for every $i \in [1,\ldots,\ell]$, all queries in $\Q^{(i)}$ are $i$-balanced with respect to the partition $\Gamma^{(\ell)}$. If this event occurs, the adversary answers all queries in $\Q^{(\ell)}$ with respect to the partition $\Gamma^{(\ell)}$, and once again, by~\Cref{lem:suff-indis}, answers to all queries in $\Q^{(1)}, \Q^{(2)}, ..., \Q^{(\ell-1)}$ remain unchanged if we answer them using the  partition $\Gamma^{(\ell)}$. The game then continues to the next round. Otherwise, with probability at most $\ell/n^3 \le 1/n^2$, the game terminates and the adversary reveals the entire partition $\Gamma^{(\ell-1)}$ to the algorithm. 

Summing up over all rounds $1$ through $s\leq \frac{r}{2}-1$, the probability that the game reaches round $s$ is at least $1 - s/n^2 \ge 1 - 1/n$. This, in turn, implies that with probability $\geq 1-\frac{1}{n}$, the random equipartition $\cP$ satisfies the following property : all the queries in $\Q^{(i)}$ are $i$-balanced with respect to $\cP$ for all $i \in [1..s]$. Now, since $s\leq \frac{r}{2}$, by~\Cref{lem:igood-indis} we get that the answers $\Ans^{(1)}, \ldots, \Ans^{(s)}$ given to these queries are the same for $f_\cP$ and $\finfunc_\cP$. Hence the algorithm cannot distinguish between these two cases. This completes the proof of~\Cref{thm:parallel-sfm-lb}.

\subsection{Modification to boost gap : $\Omega(1/\eps)$-lower bound for $\eps$-approximate SFM}\label{sec:main-thm-approx}
An inspection of the proof of~\Cref{thm:parallel-sfm-lb} shows us that 
the minimum values of $f_\cP$ and $\finfunc_\cP$ are $0$ and $-\frac{g}{2}$ for all $\cP$'s (by~\Cref{lem:unique-minimizer}). That is, any 
 polynomial query algorithm making fewer than $\tilde{\Omega}(N^{1/3})$ rounds of adaptivity cannot distinguish between the case when the minimum value is $0$ and minimum value is $-g/2$.
Since $g = \Theta(\sizeUniv^{1/3} (c \log \sizeUniv)^{2/3})$, we also rule out {\em additive} $O(\sizeUniv^{1/3})$-approximations for submodular functions whose range is $\{-N, -N+1,\ldots, N\}$. Scaling such that the range is $[-1,+1]$, we in fact obtain an $\Omgt(1/\sqrt{\eps})$-dept lower bound to obtain $\eps$-additive approximation algorithms.

In this section we show how a small modification leads to indistinguishability between functions with minimum value $0$ and those with minimum value $-\Theta(N^{2/3})$ 
thus proving an $\Omgt(\frac{1}{\eps})$ lower bound on the depth required for polynomial query $\eps$-additive approximation algorithms for SFM.

The difference is in the definition of $\finpartfunc$; we redefine it such that the minimizer is not just $P_r$ (or rather $(0,0,\ldots, 0,n)$) but $P_{\th} \cup P_{\th+1} \cup \cdots \cup P_r$, and the minimum value becomes $-\frac{gr}{6} = -\Theta(N^{2/3})$. However, it still remains indistinguishable from $\lbfunc$ if the number of rounds is $< r/2$, and thus the proof of~\Cref{thm:parallel-sfm-lb} carries word-to-word. 

\begin{mdframed}[backgroundcolor=gray!05,topline=false,bottomline=false,leftline=false,rightline=false] 
	Define $\trunc{\bx} := \left(\bx_1, \ldots, \bx_{\th -1}, \min(\bx_\th, \frac{n}{2} - \frac{g}{4}), \min(\bx_{\th+1}, \frac{n}{2} - \frac{g}{4})\ldots, \min(\bx_r, \frac{n}{2} - \frac{g}{4})\right)$. Then,
	\begin{equation}\label{eq:extension-2}
		\finprimepartfunc(\bx) = \lbfunc(\trunc{\bx}) - \sum_{i=\th}^{r} \max\left(0, \bx_i - \left(\frac{n}{2} - \frac{g}{4}\right)\right)
	\end{equation}
\end{mdframed}
\noindent
Below we note the relevant changes. Let $\finprimefunc_\cP$ be the partition submodular function induced by a partition $P = (P_1, \ldots, P_r)$ with $|P_i| = n$, and $\finprimepartfunc$.
\begin{itemize}
	\item The proof of~\Cref{lem:finpart-submod} generalizes to prove $\finprimepartfunc$ is partition submodular. The two cases are $j<  \th$ and $j \geq \th$.
	In the former case, $\partial_j\finprimepartfunc(\bx) = \partial_j\lbfunc(\trunc{\bx})$ and $\partial_j\finprimepartfunc(\by) = \partial_j\lbfunc(\trunc{\by})$, and submodularity follows
	from submodularity of $\lbfunc$. If $j \geq \th$ and $\by_j \geq \frac{n}{2} - \frac{g}{4}$, then $\partial_j\finprimepartfunc(\by) = -1$ which implies it's $\leq \partial_j\finprimepartfunc(\bx)$.
	Otherwise, both $\bx_j,\by_j < \frac{n}{2}-\frac{g}{4}$, and then submodularity again follows from that of $\lbfunc$.

	\item In~\Cref{lem:unique-minimizer}, we can now assert $\finprimefunc_\cP(P_{\th}\cup \cdots \cup P_r) = -\frac{g}{2}\cdot \frac{r}{3} =-\frac{gr}{6}$.
	\item We assert that~\Cref{lem:suff-indis} still holds. To see this, note that the only changes are in the proof of~\Cref{lem:igood-indis} (not the statement), and we sketch this below.
	Let $k := \sum_{i=\th}^{r} \max\left(0, \bx_\th - \left(\frac{n}{2} - \frac{g}{4}\right)\right)$; we (still) have $\norm{\bx}_1 = \norm{\trunc{\bx}}_1 + k$ and $\finprimepartfunc(\bx) = \lbfunc(\trunc{\bx}) - k$. Furthermore, for any $1\leq t \leq \th$, 
	we have $\ell_t(\bx) = \ell_t(\trunc{\bx}) + k$, and so if the odd-even index $\{a,b\}$ of $\bx$ is in $\{1,\ldots, \th\}$, then $\{a,b\}$ is also the odd-even index for $\trunc{\bx}$.
	
	Now, if $\bx_t\leq \frac{n}{2} - \frac{\gap}{4}$ for all $\th \leq t \leq r$, then $\trunc{\bx} = \bx$ and $k = 0$ and $\finprimepartfunc(\bx) = \lbfunc(\bx)$. So, we may assume that 
	some $\bx_t >  \frac{n}{2} - \frac{\gap}{4}$. And since $\bx$ is $i$-balanced (for $i < t$), we get (just as in the previous proof) $\bx_j \geq \frac{n}{2} - \frac{g}{2}$ for all $j\geq i$. 
	And thus, the odd-even index $\{a,b\}$ of $\bx$ lies in $\{1,2,\ldots, i+1\}$. The rest of the proof now proceeds exactly as in~\Cref{lem:igood-indis}.
\end{itemize}

\section{Suffix Functions, Nested Matroids, and Parallel Matroid Intersection}\label{sec:matroids}

In this section we explain how our suffix functions, and as a result our partition submodular functions, arise in the context of matroid intersection. 
This is then used to prove~\Cref{thm:parallel-mati-lb} which 
states that any efficient matroid intersection algorithm, even with access to {\em rank} functions to the two matroids, must proceed in polynomially many rounds.

\paragraph{Matroids.}
A matroid $\cM = (\Univ,\cI)$ is a set-system over a universe $\Univ$ satisfying the following two axioms
\begin{itemize}[noitemsep]
	\item $I\in \cI$ and $J\subseteq I$ implies $J\in \cI$.
	\item For any $I,J\in \cI$ with $|I| < |J|$, there exists $x\in J\setminus I$ such that $I+x\in \cI$.
\end{itemize}
The sets in $\cI$ are called {\em independent} sets of the matroid. A maximal independent set is called a {\em base}. It is well-known that all bases have the same cardinality.
There are two usual oracles to access matroids. The first is the {\bf \em independence oracle} which given a subset $S\subseteq \Univ$ returns whether $S$ is independent or not.
The second stronger oracle, and we assume an algorithm has access to this, is the {\bf \em rank oracle} which given a subset $S$ returns $\rank_\cM(S)$ which is the cardinality of the largest independent subset of $S$. It is well known that $\rank(S)$ is a submodular function whose marginals are in $\{0,+1\}$.

\paragraph{Nested Matroids.}
Let $\cC = \{\Univ = C_1 \supseteq C_2 \supseteq \cdots \supseteq C_r\}$ be a collection of nested subsets of the universe $\Univ$. Let each set $C_i$ have an associated non-negative integer capacity $\capa_i$. Let $\vec{\capa} = (\capa_1, \ldots, \capa_r)$ be the capacity vector. Then $(\cC, \vec{\capa})$ defines the following set family which 
is a matroid. Such matroids are called {\em nested matroids}  (see, for example,~\cite{FifeO17}) and are a special class of laminar matroids.
\begin{equation}
	\calM_{\cC} := ~\{I\subseteq \Univ ~:~ |I\cap C_t| \leq \capa_t, ~~~1\leq t\leq r\} \tag{Nested Matroids}
\end{equation}

Given the nested family $\cC$, there is an obvious associated partition  $\calP := (P_1, P_{2}, \ldots, P_r)$ of the universe $\Univ$ defined as
$P_r := C_r$, the minimal subset in $\cC$, and $P_j := C_{j}\setminus C_{j+1}$ for all $1\leq j < r$. Similarly, we define ``thresholds'' for each part of the partition $\cP$ as 
$\tau_r := \capa_r$, and $\tau_j := \capa_j - \capa_{j+1}$. We use $\vtau$ to denote the threshold vector $(\tau_1, \ldots, \tau_r)$.

Observe that these definitions are interchangeable : given $(\calP, \vtau)$ one gets the nested matroid defined by $(\cC, \vec{\capa})$, where $C_j = \bigcup_{t\geq j} P_t$ for all $1\leq j\leq r$, and $\capa_j = \sum_{t\geq j} \tau_t$.

\paragraph{Rank of a Nested Matroid.} 
Given a nested matroid $\cM$, let $\calP = (P_1, \ldots, P_r)$ be the 
associated partition with thresholds $\tau_1$ to $\tau_r$. For simplicity, let us assume $|P_i| = n$ for all $1\leq i\leq r$.
Given a subset $S\subseteq \Univ$, let $\bx\in \ZZ_{\geq 0}^r$ be the signature of $S$ where $\bx_i := |P_i\cap S|$.
Define
\begin{equation}\label{eq:lin-con-mat}
	\text{for any $1\leq t\leq r$,}~~~	\suffix_t(\bx) := \sum_{s = t}^r \left(\bx_s -\tau_s \right)  
\end{equation}
Note that a set $S$ is independent if and only if $\suffix_t(\bx) \leq 0$ for all $1\leq t\leq r$.
Also note the connection with~\eqref{eq:lin-con} when we set $\tau_1 = \cdots = \tau_{r-1} = \left(\frac{n}{2} - \gap\right)$ and $\tau_r = \left(\frac{n}{2} - \gap\right) + \gaptwo$.
The next lemma shows how these functions define the rank of a nested matroid.
\begin{mdframed}[backgroundcolor=gray!10,topline=false,bottomline=false,leftline=false,rightline=false] 	
\begin{lemma}[Rank of a Nested Matroid]\label{lem:rk-of-nested-matroid}~
	
	\noindent
	Let $\calM$ be a nested matroid defined by $(\calP = (P_1, \ldots, P_r); \vtau = (\tau_1, \ldots, \tau_r))$ where $\tau_i \geq 0$ for all $i$.
	Given any subset $S\subseteq \Univ$ with signature $\bx$, the rank of $S$ is
	\[
	\rank_\calM(S) = \norm{\bx}_1 - \max\left(0, \max_{1\leq a\leq r} \suffix_a(\bx)\right) 
	\]
	where $\suffix_t(\bx)$ is as defined in~\eqref{eq:lin-con-mat}.
\end{lemma}
\end{mdframed}
\begin{proof}
	The rank $\rank_\cM(S)$, which we also denote as $\rank_\cM(\bx)$, is the cardinality of the largest independent subset of $S$.
	This value can be found by the following linear program, which is integral because the constraint matrix is totally unimodular.
	
	\begin{minipage}{0.45\textwidth}
		\begin{alignat}{4}
			\rank(\bx) := \max & \sum_{i=1}^r \by_i \notag \\
			&\by_i \leq \bx_i,&&~~~\forall i\in [r] \notag \\
			& \sum_{i\geq t} \by_i \leq \sum_{i\geq t} \tau_i,&&~~~\forall t\in [r] \notag
		\end{alignat}
	\end{minipage} $\underbrace{=}_{\text{Duality}}$ 
	\begin{minipage}{0.45\textwidth}
		\begin{alignat}{4}
			\min & \sum_{i=1}^r \eta_i\bx_i + \sum_{t=1}^r \bz_t \cdot \left(\sum_{i\geq t}\tau_i\right) \notag \\
			& \sum_{t\leq i} \bz_t + \eta_i = 1,&&~~~\forall i\in [r] \notag \\
			& \bz, \eta \geq 0 \notag
		\end{alignat}
	\end{minipage}

	We do not impose non-negativity constraints on the $\by_i$ variables in the primal because the maximizing solution will indeed have non-negative $\by_i$'s.
	To see this, suppose $\by_j < 0$ and let $t\leq j$ be the largest index such that $\sum_{i\geq t} \by_i = \sum_{i\geq t} \tau_i$. That is, the largest indexed
	constraint, among the ones containing $\by_j$, which is tight. There must be such a $t$ for otherwise we could increase the objective by incrementing $\by_j$.
	Furthermore, $\by_t > 0$ for otherwise $\sum_{i\geq t+1} \by_i = \sum_{i\geq t+1} \tau_i$ and our $t$ won't be largest; this argument uses $\tau_t \geq 0$.
	Now, increasing $\by_j$ and decreasing $\by_t$ by the same amount gives a feasible solution with the same optimum, and continuing the above procedure, we will get to 
	a non-negative $\by$.

	We can massage the dual as follows. Let $\pref_i(\bz) := \sum_{t\leq i} \bz_t$. Thus, we can rewrite $\eta_i = 1 - \pref_i(\bz)$, and since $\eta_i\geq 0$, we get 
	all $\pref_i(\bz)$'s, and in particular which is equivalent to, by the non-negativity of $\bz$, the constraint $\norm{\bz}_1 \leq 1$.
	Therefore, we can eliminate $\eta$'s and get
	\[
	\rank(\bx) = \min_{\bz:\norm{\bz}_1\leq1} ~~\sum_{i=1}^r \bx_i \cdot \left(1 - \pref_i(\bz)\right) + \sum_{t=1}^r \bz_t  \left(\sum_{i\geq t}\tau_i\right)
	\]
	Next, using the observation that $\sum_{t=1}^r \bz_t  \left(\sum_{i\geq t}\tau_i\right) = \sum_{i=1}^r \pref_i(\bz)\cdot \tau_i$, we can further simplify to 
	get
	\[
	\rank(\bx) = \min_{\bz:\norm{\bz}_1\leq1} ~~\sum_{i=1}^r \bx_i - \sum_{i=1}^r \pref_i(\bz) \cdot \left(\bx_i - \tau_i\right) = \norm{\bx}_1 - \max_{\bz:\norm{\bz}_1\leq1} \sum_{t=1}^r \bz_t \underbrace{\left(\sum_{i\geq t} \left(\bx_i -\tau_i\right)\right)}_{\suffix_t(\bx)}
	\]
	The last summand $\max_{\bz:\norm{\bz}_1\leq1} \sum_{t=1}^r \bz_t \suffix_t(\bx)$ is $0$ if all $\suffix_t(\bx) \leq 0$ (by setting $\bz \equiv \bzero$), and otherwise, it is $\max_{1\leq a\leq t} \suffix_a(\bx)$. This completes the proof.
\end{proof}
\noindent
The reader should notice the similarity with \eqref{eq:value}. We will now make the connection more precise. Before doing so, we need another well known definition.

\paragraph{Duals of Matroids.} Given a matroid $\cM$, the dual matroid $\cM^*$ is defined as follows
\[
\calI^* := \{S\subseteq \Univ~:~ \Univ \setminus S ~~\textrm{contains a base of}~~\calM\}
\]
It is not too hard to check this is a matroid. The rank of any set in the dual matroid can be computed using 
the rank of the original matroid as follows.
\begin{lemma}[e.g., Theorem 39.3 in~\cite{Schr2003}]\label{lem:dual-rank}
	Let $\calM$ be a matroid with rank function $\rank$. Let $\calM^*$ be its dual with corresponding rank function $\rank^*$. Then,
	\[
	\forall S\subseteq \Univ:~~~\rank^*(S) = \rank(\Univ\setminus S) + |S| - \rank(\Univ)
	\]
\end{lemma}
It is not too hard to see that the dual of a nested matroid is another nested matroid whose nesting is from the ``other end''. More formally, one can prove the following.
\begin{lemma}\label{lem:dual-nested-matroid}
	Let $\cM$ be a nested matroid defined by the partition $\cP = (P_1, \ldots, P_r)$ and thresholds $\vtau := (\tau_1, \ldots, \tau_r)$.
	Then, $\cM^*$ is another nested matroid defined by the {\em reverse} partition $\cP' = (P_r, P_{r\!-1}, \!\!\ldots, \!\!P_2, P_1)$ 
	and thresholds $\vtau' := (n_r - \tau_r, n_{r-1} - \tau_{r-1}, \ldots, n_1 - \tau_1)$, where $n_i := |P_i|$.
\end{lemma}
\begin{proof}
	Let $S$ be a subset with signature $\bx$ with respect to the original partition $\cP$. $S$ is independent in $\cM^*$ if and only if $\Univ \setminus S$ contains a base of $\cM$.
	Equivalently, $\rank_\cM(\Univ\setminus S) = \rank_\cM(U)$. Now, the latter is precisely $\norm{\bn}_1 - \ell_1(\bn)$ where $\bn = (n_1, n_2, \ldots, n_r)$ is the signature of the universe $\Univ$. Let $\bz$ be the signature of $\Univ\setminus S$; note that $\bz_i = n_i - \bx_i$. 
	Thus, we get that $S$ is independent in $\cM^*$ if and only if
	\[
\norm{\bz}_1 - \max(0, \max_{1\leq a\leq r} \ell_a(\bz)) = \norm{\bn}_1 - \ell_1(\bn) ~~\underbrace{\Rightarrow}_{\text{Rearranging}} \suffix_1(\bz) = \max(0, \max_{1\leq a\leq r} \suffix_{a}(\bz))
	\]
$\ell_1(\bz)$ is largest suffix if and only if all the $(r-1)$ {\em prefix-sums} are non-negative, and $\ell_1(\bz) \geq 0$ implies all prefix-sums are non-negative.
Thus, we get 
\[
\forall 1\leq j\leq r, ~~\sum_{j\leq t} (\bz_j - \tau_j) \geq 0  ~~~\equiv~~~ \forall 1\leq j\leq r, ~~\sum_{j\leq t} \left(\bx_j - (n_j -\tau_j)\right) \leq 0
\]
which is precisely the signature of an independent set in the nested matroid defined by $(\cP',\vtau')$.
\end{proof}

\paragraph{The Hard Matroid Intersection Set-up.} Let $r = 2k+1$ be an odd number. Let $\calP = (P_1, \ldots, P_r)$ be a partition with $|P_i| = n$. Each part will be associated with a parameter $\tau_i$.
These will be set to $\tau_1 = \cdots = \tau_{r-1} = \left(\frac{n}{2} - \gap\right)$ and $\tau_r = \left(\frac{n}{2} - \gap\right) + \gaptwo$, where $\gap,\gaptwo$ are as described in~\Cref{sec:main-thm}.

We define {\em three} coarsenings of this partition. The first is the {\em odd} coarsening containing $(k+1)$ parts defined as follows.
\[
\calPodd := \left(P_1\cup P_2, ~~P_3 \cup P_4, ~~\ldots, ~~P_{r-2} \cup P_{r-1}, ~~P_r\right)
\]
and the associated $\tau$-values are, as expected, the sum of the relevant $\tau_j$'s. More precisely, they are $\vtau_{\mathsf{odd}} := (\tau_1 + \tau_2, \tau_3 + \tau_4, \ldots, \tau_{r-2}+\tau_{r-1}, \tau_r)$.
Let $\calModd$ be the nested matroid defined by $(\calPodd,\vtau_{\mathsf{odd}})$. The rank of $\calModd$ is given by~\Cref{lem:rk-of-nested-matroid} as follows; we only consider the odd indices since $r$ is odd.

\begin{claim}\label{clm:rk-modd}
	Let $S\subseteq \Univ$. Let $\bx$ be the signature of $S$ with respect to the $(2k+1)$-part partition $\calP$.
	Then, \[\rank_{\calModd}(\bx) := \rank_{\calModd}(S) = \norm{\bx}_1 - \max\left(0, \max_{1\leq a\leq r, ~a~\textrm{odd}} \suffix_a(\bx)\right)\]
\end{claim}
\noindent
The second coarsening is the {\em even} coarsening containing $(k+1)$-parts defined as 
\[
\calPeven := \left(P_1, ~~ P_2\cup P_3, ~~P_4 \cup P_5, ~~\ldots, ~~P_{r-1} \cup P_{r}\right)
\]
The associated $\tau$-values are slightly different in that the first part is effectively ``ignored''. The vector of $\tau$'s are $\vtau_{\mathsf{even}} := (n, \tau_2 + \tau_3, \tau_4 + \tau_5, \ldots, \tau_{r-1}+\tau_r)$.
Let $\calMeven$ be the corresponding nested matroid defined by $(\calPeven, \vtau_{\mathsf{even}})$. Note that any base of $\calMeven$ must contain the whole set $P_1$. Again using ~\Cref{lem:rk-of-nested-matroid}, the rank of this matroid is given as follows. 

\begin{claim}\label{clm:rk-meven}
	Let $S\subseteq \Univ$. Let $\bx$ be the signature of $S$ with respect to the $(2k+1)$-part partition $\calP$.
	Then, \[\rank_{\calMeven}(\bx) := \rank_{\calMeven}(S) = \norm{\bx}_1 - \max\left(0, \max_{1\leq a\leq r, ~a~\textrm{even}} \suffix_a(\bx)\right)\]
\end{claim}

The reason the first part does not count is because $(\bx_1 - n)$ is $\leq 0$, and this cannot be the maximizer when we apply \Cref{lem:rk-of-nested-matroid}. And otherwise, it corresponds to an even index in the original partition. \smallskip

Finally, the third coarsening is a refinement of $\calPeven$ where the last part $P_{r-1} \cup P_r$ is divided into two.
That is,
\[
\calPeven' := \left(P_1, ~~ P_2\cup P_3, ~~P_4 \cup P_5, ~~\ldots, ~~P_{r-1},~~  P_{r}\right)
\]
The associated $\tau$ vector is $\vtau'_{\mathsf{even}} := (n, \tau_2 + \tau_3, \tau_4 + \tau_5, \ldots, \tau_{r-1} + \tau_r - \theta, \theta)$ for some parameter $\theta$, which is set to $\left(\frac{n}{2} - \frac{g}{4}\right)$.
Let $\calMeven'$ be the nested matroid defined by $(\calPeven',\vtau'_{\mathsf{even}})$.

\begin{claim}\label{clm:rk-mevenprime}
	Let $S\subseteq \Univ$. Let $\bx$ be the signature of $S$ with respect to the $(2k+1)$-part partition $\calP$.
	Then, \[\rank_{\calMeven'}(\bx) := \rank_{\calMeven'}(S) = \rank_{\calMeven}(\trunc{\bx})\]
	where, $\trunc{\bx} = \left(\bx_1, \ldots, \bx_{r-1}, \min(\bx_r, \theta)\right)$.
\end{claim}
\begin{proof}
	First observe that for any $t$, $\suffix_t(\bx) = \suffix_t(\trunc{\bx}) + \max(0,(\bx_r - \theta))$. Therefore, for any $\bx$, the $t$ maximizing $\ell_t(\bx)$ also is the one maximizing $\ell_t(\trunc{\bx})$.

	When computing $\rank_{\calMeven'}(\bx)$ as $\norm{\bx} - \max(0,\max_a \suffix_a(\bx))$, the maximization over $a$ is over all even indices and also $r$. This leads to two cases.
	
	Case 1: This maximizer is at $a = r$, that is, $\rank_{\calMeven'}(\bx) = \norm{\bx}_1 - \max(0, \bx_r - \theta)$. In that case, we have $\suffix_a(\bx) \leq (\bx_r - \theta)$ for all other $a$'s. Which implies $\suffix_a(\trunc{\bx}) \leq 0$.
	Therefore, $\rank_{\calMeven}(\trunc{\bx}) = \norm{\trunc{\bx}}_1 = \norm{\bx}_1 - \max(0, \bx_r - \theta) = \rank_{\calMeven'}(\bx)$.
	
	Case 2: This maximizer at $a\neq r$, that is, $\rank_{\calMeven'}(\bx) = \norm{\bx}_1 - \max(0,\suffix_a(\bx))$ for some even $a$.
	Note that this $a$ is also the maximizer when computing $\rank_{\calMeven}(\trunc{\bx})$. Therefore,
	\[\rank_{\calMeven}(\trunc{\bx}) = \norm{\trunc{\bx}}_1 - \max(0,\suffix_a(\trunc{\bx})) = \norm{\trunc{\bx}}_1 - \max(0,\suffix_a(\bx) - \max(0,\underbrace{(\bx_r - \theta)}_{\suffix_r(\bx)}))\]
	
	If $\bx_r \leq \theta$, we get $\rank_{\calMeven}(\trunc{\bx}) = \norm{\trunc{\bx}}_1 - \max(0,\suffix_a(\bx)) = \norm{\bx}_1 - \max(0,\suffix_a(\bx)) = \rank_{\calMeven'}(\bx)$,
	where the second equality follows because $\trunc{\bx} = \bx$ when $\bx_r \leq \theta$.
	
	If $\bx_r > \theta$, then $\rank_{\calMeven}(\trunc{\bx}) = \norm{\trunc{\bx}}_1 - \left(\suffix_a(\bx) - (\bx_r - \theta)\right)$ since $\suffix_a(\bx) \geq \suffix_r(\bx) \geq 0$ as $a$ is the maximizer. Now observe that $\norm{\trunc{\bx}}_1 = \norm{\bx}_1 - \left(\bx_r -\theta\right)$, and so $\rank_{\calMeven}(\trunc{\bx}) = \norm{\bx} - \suffix_a(\bx) = \rank_{\calMeven'}(\bx)$.
\end{proof}

\begin{claim}
	$\rank_{\calMeven}(\Univ) = \rank_{\calMeven'}(\Univ)$.
\end{claim}
\begin{proof}
	Let $\bn$ be the $(n,n,\ldots, n)$ vector.
	$\rank_{\calMeven}(\Univ) = \norm{\bn}_1 - \suffix_2(\bn)$, and $\rank_{\calMeven'}(\Univ) = \rank_{\calMeven}(\trunc{\bn})$.
	This, in turn, is $\norm{\trunc{\bn}}_1 - \suffix_2(\trunc{\bn}) = (\norm{\bn} - (n - \theta)) - \left(\suffix_2(\bn) - (n-\theta)\right) = \norm{\bn}_1 - \suffix_2(\bn)$.
\end{proof}

The following lemma connects matroid intersection with submodular function minimization for the functions described in~\Cref{sec:formal}.
\begin{mdframed}[backgroundcolor=gray!10,topline=false,bottomline=false,leftline=false,rightline=false] 	
\begin{lemma}\label{lem:mat-dual-two}
	The size of the largest cardinality independent set in $\calModd \cap \calMeven^*$ is precisely $C + \min_{S\subseteq \Univ} f(S)$ where
	$C = |\Univ| - \rank_{\calMeven}(\Univ)$ and $f(S) = \lbfunc(\bx)$ with 
	\[
	\lbfunc(\bx) = \norm{\bx}_1 - \max\left(0, \max_{1\leq a\leq r, ~a~\textrm{odd}} \suffix_a(\bx)\right) - \max\left(0, \max_{1\leq a\leq r, ~a~\textrm{even}} \suffix_a(\bx)\right)
	\]
	and the size of the largest cardinality independent set in $\calModd \cap (\calMeven')^*$ is precisely $C + \min_{S\subseteq \Univ} f^*(S)$ where
	$C = |\Univ| - \rank_{\calMeven'}(\Univ) = |\Univ| - \rank_{\calMeven}(\Univ)$ and $f^*(S) = \finpartfunc(\bx)$ with 
	\[
	\finpartfunc(\bx) = \begin{cases}
		\lbfunc(\bx) & \textrm{if}~ \bx_r \leq \theta \\
		\lbfunc(\trunc{\bx}) - \left(\bx_r - \theta\right) & \textrm{otherwise}
	\end{cases}
	~~\textrm{where}, \trunc{\bx} := \left(\bx_1, \ldots, \bx_{r-1}, \min(\bx_r, \theta)\right)
	\]
\end{lemma}
\end{mdframed}
\begin{proof}
	From Edmond's theorem~\cite{Edmo1970}, we know that for any two matroids $\cM_1$ and $\cM_2$, one has
	\[
	\max_{I \in \cM_1 \cap \cM_2} |I| ~=~ \min_{S\subseteq \Univ} \left(\rank_{\cM_1} (S)  + \rank_{\cM_2}(\Univ\setminus S)\right)
	\]
	Fix a set $S$ with signature $\bx$ with respect to the $(2k+1)$-part partition $\cP$. 
By ~\Cref{clm:rk-modd}, we have $\rank_{\calModd}(S) = \rank_{\calModd}(\bx) = \norm{\bx}_1 -  \max\left(0, \max_{1\leq a\leq r, ~a~\textrm{odd}} \suffix_a(\bx)\right)$.
By ~\Cref{lem:dual-rank}, we have $\rank_{\calMeven^*}(U\setminus S) = \rank_{\calMeven}(S) + |U| - \rank_{\calMeven}(U) - |S| = \rank_{\calMeven}(\bx) + C - \norm{\bx}_1 $. 		
By ~\Cref{clm:rk-meven}, we have $\rank_{\calMeven}(S) = \rank_{\calMeven}(\bx) = \norm{\bx}_1 -  \max\left(0, \max_{1\leq a\leq r, ~a~\textrm{even}} \suffix_a(\bx)\right)$.
And thus, 
\[
\rank_{\calModd}(S) + \rank_{\calMeven^*}(\Univ\setminus S) = C + h(\bx)
\]
Similarly, by~\Cref{lem:dual-rank}, we have $\rank_{\left(\calMeven'\right)^*}(U\setminus S) = \rank_{\calMeven'}(S) + |U| - \rank_{\calMeven}(U) - |S| = \rank_{\calMeven'}(\bx) + C - \norm{\bx}_1 $.
By~\Cref{clm:rk-mevenprime}, the RHS equals $\rank_{\calMeven}(\trunc{\bx}) + C - \norm{\bx}_1$.
And so, 
\[
\rank_{\calModd}(S) + \rank_{\left(\calMeven'\right)^*}(\Univ\setminus S) = C + \norm{\trunc{\bx}}_1  -  \max\left(0, \max_{1\leq a\leq r, ~a~\textrm{odd}} \suffix_a(\bx)\right)  - \max\left(0, \max_{1\leq a\leq r, ~a~\textrm{even}} \suffix_a(\trunc{\bx})\right) 
\]
When $\bx_r \leq \theta$, the RHS is $C + h(\bx)$. When $\bx_r > \theta$, we have $\ell_t(\bx) = \ell_t(\trunc{\bx}) + (\bx_r - \theta)$ for all $t$, and as before, one can argue that
$\max\left(0, \max_{1\leq a\leq r, ~a~\textrm{odd}} \suffix_a(\bx)\right) = \max\left(0, \max_{1\leq a\leq r, ~a~\textrm{odd}} \suffix_a(\trunc{\bx})\right) + (\bx_r - \theta)$.
Which implies the RHS is $C + h(\trunc{\bx}) - (\bx_r -\theta)$. In sum, the RHS is $C + h^*(\bx)$.
\end{proof}
\paragraph{An Illustration.}
It is perhaps instructive to illustrate the difference in the two situations described in \Cref{lem:mat-dual-two} with a concrete example which directly
describes why the largest cardinality common independent sets are different in the two different cases.
Take $r=3$.
Fix a partition $(P_1,P_2,P_3)$ with each part having $n$ elements each, and the size of the universe is $3n$. The $\tau$ values are $(\frac{n}{2} - g, \frac{n}{2} - g, \frac{n}{2} - 0.25g)$.

Let us understand what $\calModd$ is in this case. This is generated by $(P_1 \cup P_2, P_3)$ and the threshold vector $(n - 2g, \frac{n}{2} - 0.25g)$. So, a subset $I$ is independent in $\calModd$ iff (a) it contains $\leq \frac{n}{2} - 0.25g$ elements from $P_3$, and (b) $\leq \frac{3n}{2} - 2.25g$ elements overall.

Similarly, the matroid $\calMeven$ is generated by $(P_1, P_2 \cup P_3)$ with the threshold vector $(n, n - 1.25g)$. We are interested in its dual, which is also a nested matroid which, by~\Cref{lem:dual-nested-matroid} is generated by the partition $(P_2\cup P_3, P_1)$ with thresholds $(n + 1.25g, 0)$. That is, a subset $I$ is independent in $\calMeven^*$ iff
(a) it contains $0$ elements from $P_1$, and (b) $\leq n+1.25g$ elements overall. 

Notice that any set $I^*$ which contains $\frac{n}{2} - 0.25g$ elements from $P_3$, $\frac{n}{2} + 1.5g$ elements from $P_2$, and $0$ elements from $P_1$ is a {\bf \em base} of $\calMeven$ which is independent in $\calModd$. All that is needed is that $1.5g \leq \frac{n}{2}$ so that there are enough items in $P_2$ to pick from.

Finally, let us consider the matroid $\left(\calMeven'\right)$ and its dual. The former is a nested matroid generated by $(P_1, P_2, P_3)$ with thresholds 
$(n, \frac{n}{2} - g, \frac{n}{2} - 0.25g)$. Which, in turn, implies that its dual is a nested matroid generated by $(P_3, P_2, P_1)$ with thresholds 
$(\frac{n}{2} + 0.25g, \frac{n}{2} + g, 0)$. That is, an independent set cannot contain more than $\frac{n}{2} + g$ elements from $P_2$, thus ruling out the $I^*$ described in the previous paragraph. Indeed, since $\calModd$ forces at most $\frac{n}{2} - 0.25g$ elements from $P_3$, the largest common independent set in $\calModd$ and $(\calMeven')^*$ is 
at most of size $n + 0.75g$ elements. Which is exactly $-g/2$ less, as predicted by~\Cref{lem:mat-dual-two} and~\Cref{lem:unique-minimizer}.
Note, however, that the size of the largest independent set in $(\calMeven')^*$ is the same as that in $\calMeven^*$, that is $n + 2.75g$; that set picks more elements from $P_3$. It is the intersection with $\calModd$ which prevents picking such a base of $(\calMeven')^*$.

\begin{proof}[Proof of~\Cref{thm:parallel-mati-lb}]
To complete the proof of~\Cref{thm:parallel-mati-lb}, we need one more thing. In SFM, we have access to evaluation oracle for the function. In particular, if $\bx$ is the signature of a set $S$ with respect to a partition, then we have access to $\lbfunc(\bx)$. In the matroid intersection problem, we have access to the individual ranks of each matroid. Therefore, we need to establish suffix-indistinguishability for each of the individual ranks. Since the rank of the dual matroid can be simulated by the rank of the original matroid, the suffix indistinguishability of both matroids is established by the following lemma whose proof is very similar to that of~\Cref{lem:suff-indis}. 

\begin{lemma}\label{lem:suff-indis-matroids}
	A signature $\bx$ (with respect to the original $(2k+1)$-part partition) is $i$-balanced if 
	$\bx_i - \frac{g}{8} \leq \bx_j \leq \bx_i + \frac{g}{8}$.
	Let $i < \frac{r}{2}$.
	If $\bx$ and $\bx'$ are two $i$-balanced points
	with $\bx_j = \bx'_j$ for $j\leq i$ and $\norm{\bx}_1 = \norm{\bx'}_1$, then 
	(a) $\rank_{\calModd}(\bx) = \rank_{\calModd}(\bx')$, and 
	(b) $\rank_{\calMeven}(\bx) = \rank_{\calMeven}(\bx') = \rank_{\calMeven'}(\bx) = \rank_{\calMeven'}(\bx')$
\end{lemma}
\begin{proof}
	As in the proof of~\Cref{lem:suff-indis}, we proceed in two claims.
	First, we claim that for any $i \leq r-2$, if $\bx$ and $\bx'$ are $i$-balanced, then $\rank_{\calMeven}(\bx) = \rank_{\calMeven}(\bx')$.
	If $\bx_i = \bx'_i < \frac{n}{2} - \frac{7g}{8}$, then just as in~\Cref{clm:suff-indis-lbfuc}, all $\bx_j,\bx'_j$, for $j\geq i$, are $\leq \frac{n}{2}-\frac{3g}{4}$, implying that the even-index
	with the largest $\ell_t(\cdot)$ must lie in $\{1,2,\ldots, i+1\}$. And this, due to the premise of the lemma, implies (using~\Cref{clm:rk-meven}) $\rank_{\calMeven}(\bx) = \rank_{\calMeven}(\bx')$.
	A similar argument using odd-index and~\Cref{clm:rk-modd} proves part (a).
	
	The proof of the second and third equality in part(b) follows as in~\Cref{lem:igood-indis}. We have $\theta = \frac{n}{2} - \frac{g}{4}$. If $\bx_r \leq \theta$, then the two ranks are the same by~\Cref{clm:rk-mevenprime}. If $\bx_r > \theta$, then since $\bx$ is $i$-balanced, all $\bx_j \geq \frac{n}{2}-\frac{g}{2}$ for $j\geq i$. This means the even index with the largest $\ell_t(\bx)$
	lies in $\{1, \ldots, i+1\}$. And since $i\leq r/2$, which implies that both $\ell_i(\trunc{\bx})$ and $\ell_{i+1}(\trunc{\bx})$ (we look at both for we don't know which is even, but one of them is) are $\geq 0$. Therefore, $\rank_{\calMeven}(\bx) = \norm{\bx}_1 - \ell_a(\bx)$ for some even $a \leq i+1$, and $\rank_{\calMeven'}(\bx) = \norm{\trunc{\bx}}_1 - \ell_a(\trunc{\bx})$ for the same $a$.
	Since $a\leq i+1$, we get that $\norm{\trunc{\bx}}_1 = \norm{\bx}_1 - k$ and $\ell_a(\trunc{\bx}) = \ell_a(\bx) - k$, where $k = \bx_r - \theta$. In sum, we get $\rank_{\calMeven}(\bx) = \rank_{\calMeven'}(\bx)$, and this, together with the previous paragraph, implies part (b).
\end{proof}
The proof of~\Cref{thm:parallel-mati-lb} then follows almost word-to-word as the proof of~\Cref{thm:parallel-sfm-lb}. The hard distributions over the pairs of matroids are as follows.
First one samples a random equipartition $P$ of $\Univ$ into $(2k+1)$ parts. Given $P$, the ``odd'' matroid $\calModd$ is one nested matroid. The other nested matroid is 
either $\calMeven^*$ or $(\calMeven')^*$. Note that by~\Cref{lem:dual-nested-matroid}, these duals are also nested matroids. We give the algorithm rank-oracle access to these two matroids.
As in the proof of~\Cref{thm:parallel-sfm-lb}, armed with~\Cref{lem:suff-indis-matroids}, one can show that for any $s$-round deterministic algorithm for $s\leq \frac{r}{2} - 1$, with probability $\geq 1 - \frac{1}{n}$, the answers given in the case of $(\calModd, \calMeven^*)$ and the answers given in the case of $(\calModd, (\calMeven')^*)$ are exactly the same.
Since the {\em sizes} of the largest common independent sets in both cases are different, one gets the proof of~\Cref{thm:parallel-mati-lb}.
\end{proof}

\section{Concluding Remarks}

The main finding of this paper is that submodular function minimization and matroid intersection, two fundamental discrete optimization problems which have efficient algorithms, are not highly parallelizable in the oracle model. More precisely, if the access to the submodular function is via an evaluation oracle, or if the access to the matroids is via  rank oracles, then any, possibly randomized, algorithm making at most $\poly(N)$ queries to these oracles must proceed in $\tilde{\Omega}(N^{1/3})$ rounds, where $N$ is the number of elements in the universe the functions/matroids are defined on. It is an interesting question if the lower bound can be improved to $\tilde{\Omega}(N)$, or if there can be $o(N)$-round $\poly(N)$ query algorithms for either of these problems. As remarked in~\Cref{sec:technical-overview}, our constructions have a bottleneck at $N^{1/3}$, and a new idea is needed if one wants to prove better lower bounds. 

Figuring out what the query complexity of SFM and matroid intersection, regardless of adaptivity, is an intriguing question. Currently, the best known upper bounds on the query complexity for SFM is $\tilde{O}(N^2)$ ~\cite{Jiang2021}. For matroid intersection, the best known upper bounds using rank-oracles is
$\tilde{O}(N^{1.5})$ ~\cite{ChakrLSSW2019} and with
independence oracles it is $\tilde{O}(N^{9/5})$~\cite{BlikBMN2021}. For both SFM and matroid intersection, even with independence oracles, 
the best known lower bounds~\cite{GraurPRW2020,Harve2008} are only linear in $N$. 
At this juncture we mention that the submodular functions we construct in this paper can indeed be minimized in $O(N)$ queries. The main idea is that in the $i$th round one can find $O(g)$ elements from the $i$th part making only $O(gr)$ queries, and once these $O(g)$ elements are known, one can repeat the same for the next round. This gives an $O(gr^2) = O(N)$ query algorithm. Similar ideas also give $\tilde{O}(\sizeUniv)$-independence query algorithms for the matroid intersection problem, for the matroids we consider. One would need more ideas to obtain a super-linear lower bound.

We believe that {\em partition} submodular functions deserve further study in their own right. When $r=\sizeUniv$ they encompass every submodular function, and at the other extreme when $r=1$, they contain the functions which are concave functions of the cardinality. In this sense, the number of parts $r$ behaves as a measure of complexity of such functions. Can $r$-partition submodular functions be minimized in $O(\sizeUniv + \poly(r))$ or $O(\sizeUniv\cdot \poly(r))$ queries? Can partition submodular functions be minimized in $\poly(r)$ rounds, independent of $\sizeUniv$? 
We believe these questions are  worthy of study.

\bibliographystyle{alpha}
\bibliography{masterbib}
\end{document}